\documentclass[a4paper,11pt]{article}
\usepackage{tikz}
\usepackage[compat=1.1.0]{tikz-feynman}
\usepackage{jheppub} 
\usepackage{verbatim}
\usepackage[T1]{fontenc} 
\usepackage{amsmath}
\usepackage{braket}
\usepackage{float}
\usepackage{slashed}
\usepackage{graphicx} 
\usepackage{tabularx}
\usepackage{booktabs} 
\usepackage{braket}
\usepackage{mathtools}
\usepackage{caption,subcaption}
\usepackage{axodraw4j}
\usepackage{pstricks}
\usepackage{color}
\usepackage{braket}
\usepackage{cleveref}
\usepackage{multirow}
\usepackage{breqn}
\usepackage{listings}
\usepackage{autobreak}
\allowdisplaybreaks[1]

\newcommand{\ep}{\varepsilon}

\title{\boldmath Massive Two-Loop Heavy Particle Diagrams}

\author[a]{B. Assi,}
\author[a]{B.A. Kniehl,}
\author[b,c,d]{A.I. Onishchenko}

\affiliation[a]{II. Institut f\"{u}r Theoretische Physik, Universit\"{a}t Hamburg, 22761 Hamburg, Germany}
\affiliation[b]{Bogoliubov Laboratory of Theoretical Physics, Joint Institute for Nuclear Research, Dubna, Russia}
\affiliation[c]{Budker Institute of Nuclear Physics, Novosibirsk, Russia}
\affiliation[d]{Skobeltsyn Institute of Nuclear Physics, Moscow State University, Moscow, Russia,}

\abstract{We determine the master integrals for vertex and propagator diagrams that appear in effective field theories containing heavy fields. The integrals involve at least one heavy line, and the standard lines include an arbitrary mass scale. The evaluation is done analytically with modern techniques. We employ the methods of differential equations and dimensional recurrence relations to evaluate said integrals up to two-loop order.  }
\begin{document} 
\maketitle
\flushbottom
\newpage
\section{Introduction}
Heavy Particle Effective Theories (HPET) have a wide range of applicability, and its use cases have expanded dramatically in recent years. They are apparent when a field of arbitrary spin in a given theory is taken to have a large mass compared to other propagating massive degrees of freedom. HPETs were originally conceived in the context of Quantum Electrodynamics (QED) and Quantum Chromodynamics (QCD), such as in Heavy Quark Effective Theory (HQET), Non-Relativistic (NR) QCD and QED and variations therein~\cite{pineda1997effective,brambilla2000potential,georgi1990effective}. More recently, it has also been applied in the Electroweak (EW) regime~\cite{jantzen2005two,jantzen2006two, chiu2008electroweak,chiu2008electroweak0,chiu2008electroweak2,assi1,assi3}, as well as beyond the Standard Model (SM) such as in the context of heavy dark matter~\cite{ovanesyan2015heavy,beneke2019wino}, $Z'$ bosons~\cite{bauer2020low,mecaj2020effective} and black hole interactions~\cite{damgaard2019heavy,aoude2020shell,bern2021gravitational}.

When dealing with such theories beyond leading perturbative order, one is faced with loop diagrams containing eikonal lines. In this work, we determine these at two-loop order employing a set of modern techniques, in particular differential equations and dimensional recurrence relations, which have been successful in similar contexts~\cite{smirnov2010three,lee2016evaluating}. We further include a non-zero mass-scale in the standard lines for theoretical models with massive propagating degrees of freedom. The mass scale bounds the infrared (IR) regime for the two- and three-point diagrams studied here.  Even in theories with exclusively massless propagating degrees of freedom such as QED/QCD and gravity, the IR structure needs to be correctly understood~\cite{ablinger2018heavy,bern2021gravitational}. The diagrams considered here are especially useful in the evaluation of form factors of a given model. The form factor is most well-known for its uses in perturbative analyses of scattering processes occurring at the LHC and future colliders~\cite{chiu2009factorization,chiesa2013electroweak}. Form factors are of primary consideration instead of specific processes as they form the fundamental building blocks for a vast array of processes. For instance, they have been employed to study di-jet, $\bar{t}t$, squark pair, and dark matter (DM) production in various models~\cite{chiu2008electroweak2,beenakker2010supersymmetric,ovanesyan2015heavy,ciafaloni2011weak}. It is also the simplest amplitude that can be used to study the IR behaviour of a theory of interest. For further reference in the context of the SM, the QCD form factors of quarks have been evaluated to three-loop order~\cite{gehrmann2010calculation, von2017quark,bernreuther2005two,blumlein2019heavy,ablinger2018heavy}, and the EW corrections using both EFT and IR evolution equations are currently being studied to two-loop order~\cite{ ciafaloni2000electroweak,fadin2000resummation,kuhn2000summing,feucht2004two,jantzen2005two,denner2001one,hori2000electroweak,jantzen2006two,assi3}.

On the other hand, there has also been significant progress in the realm of Feynman diagram evaluation. When previously, certain classes of multi-loop diagrams were intractable, they have now become determinable with the help of novel techniques. Most notably, diagrams with masses are now attainable with the differential equations method~\cite{diffeqn1,diffeqn2,diffeqn3,diffeqn4,diffeqn5}. The basis of which is set upon differentiating the master integrals (MIs) of interest, forming a system of differential equations and reducing said system to so-called $\ep$-form~\cite{epform1,epform2,epform-criterium}. Given that such a reduction is achievable~\cite{epform-criterium} and the obtained differential system is rational, the MIs are expressible in terms of multiple polylogarithms (MPLs)~\cite{goncharov2011multiple,remiddi2000harmonic}. At present, we have also good progress in understanding equations not reducible to $\ep$-form\footnote{See for example \cite{epform-elliptics} for non-algebraic transformation to $\ep$-form in the elliptic case  and \cite{ep-regular-basis} for a notion of regular basis for non-polylogarithmic integrals.} and functions beyond multiple polylogarithms such as for example elliptical polylogarithms (EPLs)~\cite{levin1994elliptic,brown2013multiple,Adams:2014vja,bloch2015elliptic,adams2016kite,remiddi2017elliptic,broedel2018elliptic,Broedel:2018iwv,Broedel:2018qkq,Broedel:2019tlz,weinzierl2021modular, bezuglov2021massive}, or entirely novel functions~\cite{bloch2015feynman,Primo:2017ipr,adams2018planar,adams2018analytic,Bourjaily:2017bsb,Bourjaily:2018ycu,Bourjaily:2018yfy,bezuglov2021massive,Bonisch:2021yfw}. 
In our case, the diagrams we encounter are reducible to $\ep$-form with rational kernels and thus, can be written in terms of MPLs. However, when we take external lines off-shell, the integrals are only reducible to $A+B\ep$ form, as we will see. One must thus resort to EPLs to solve such MIs. This work provides results for the massive heavy-heavy, heavy-light and propagator diagrams at two-loop order. The results are explicitly given up to $\mathcal{O}(\ep^2)$ working in $D=4-2\ep$ dimensions, which is the appropriate order for SM-like theories~\cite{chiu2008electroweak,chiu2008electroweak0,chiu2008electroweak2,assi3}. However, this order is arbitrary, as the results are simply attainable for any order in $\ep$.

Outlining this paper, we begin in Section~\ref{scn:formalism} by presenting our two-loop integral families and their associated differential equations. We then reduce each set of differential equations to either $\ep$-form or $A+B\ep$ form, illustrating the sequence of balance transformations required. In Section~\ref{scn:results} we proceed to solve the differential equations and explicitly present results for each integral expanded up to an appropriate order in $\ep$. In Appendix~\ref{oneloop-masters} we further calculate the corresponding one-loop integrals for completeness.

\section{Integral families and differential equations}
\label{scn:formalism}
\begin{figure}[tb]
	\begin{center}
		\includegraphics[width=15cm]{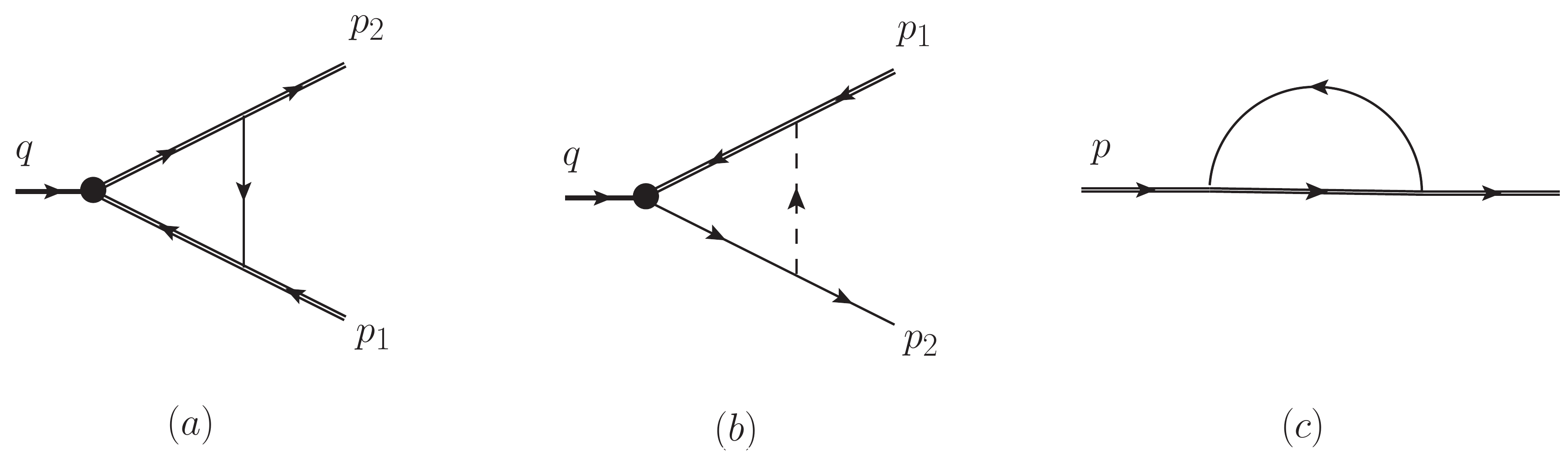}		
		\caption{Prototype topologies of one-loop vertex and self-energy diagrams: $(a)$ is heavy-heavy vertex, $(b)$ is heavy-light and $c$ is selfnergy. Solid lines represent massive particles, double lines represent heavy particles and dashed lines correspond to massless propagators. Arrows represent direction of momenta.}
		\label{fig:oneloop}
	\end{center}
\end{figure}
\begin{figure}[tb]
	\begin{center}
		\includegraphics[width=15cm]{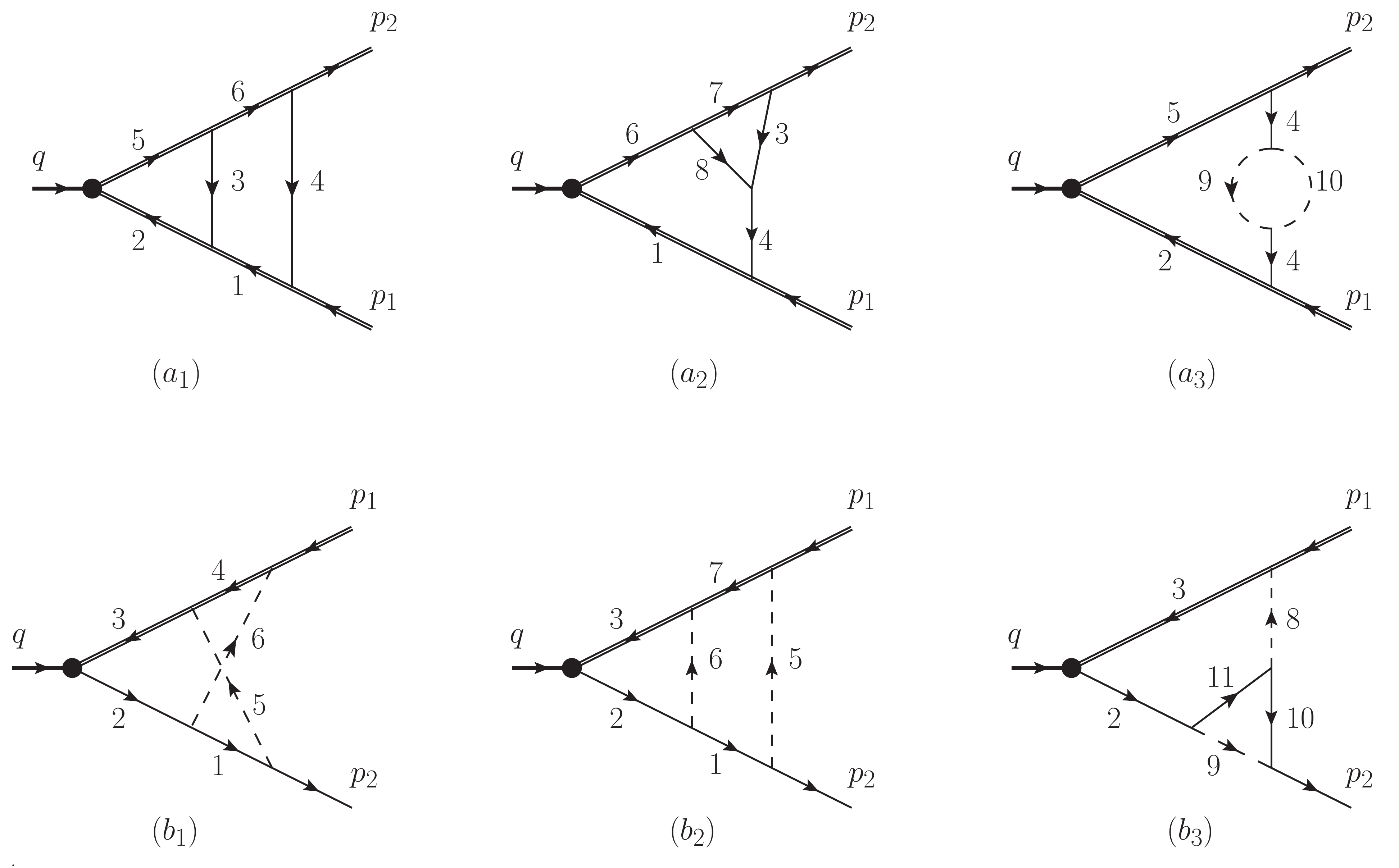}		
		\caption{Prototype topologies of two-loop vertex diagrams. Solid lines represent massive particles, double lines represent heavy particles, dashed lines correspond to massless propagators. Arrows represent direction of momenta. $(a_i)$ and $(b_i)$ correspond to heavy-heavy and heavy-light topologies. We also include the case of light self-energy insertions as is apparent in $(a_3)$.}
		\label{fig:top1}
	\end{center}
\end{figure}

\begin{figure}[tb]
	\begin{center}
		\includegraphics[width=15cm]{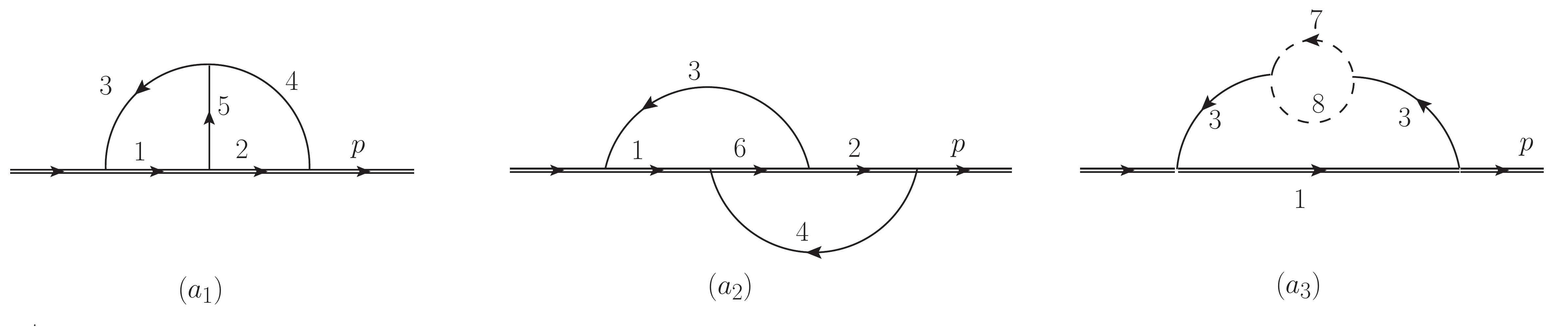}		
		\caption{Heavy field self-energy topologies. The MIs associated to other topologies are subsets of the MIs required for topologies illustrated.}
		\label{fig:top3}
	\end{center}
\end{figure}

The master integrals for one loop HPET vertices and self-energy shown in Fig.~\ref{fig:oneloop} are easy to calculate. Nevertheless, for completeness, we present their calculation in Appendix \ref{oneloop-masters}. The massive HPET vertices and self-energy at two-loop level have MIs with topologies represented by Figs.~\ref{fig:top1} $(a,b)$ and~\ref{fig:top3}, respectively. We begin by considering the  prototype topologies for the heavy-heavy vertex in Figs.~\ref{fig:top1} $(a)$, the master integrals of which can be expressed in terms of a single integral family
\begin{equation}
J^{HH}_{\nu_1,\ldots, \nu_{10}} = \int \frac{d^d l_1 d^d l_2}{(i\pi^{d/2})^2}\prod_{i=1}^{10}\frac{1}{(D_i+i0)^{\nu_i}}\, , \label{JHH-prototype}
\end{equation}
where
\begin{align}
& \quad D_1=l_2\cdot v_1, \quad D_2=l_1\cdot v_1, \quad D_3=(l_1-l_2)^2-M^2,\quad  D_4=l_2^2-M^2, \quad D_5=l_1\cdot v_2,  \nonumber \\& \quad D_6=l_2\cdot v_2, \quad D_7=(l_2-l_1)\cdot v_2, \quad D_8=l_1^2-M^2, \quad D_{9}=l_2^2,  \quad D_{10}=(l_1-l_2)^2.
\end{align}
Here, $v_{1,2}$ are the heavy particle velocities, and $M$ is the mass of exchanged bosons. It is convenient to rescale integration momenta with respect to $M$ and factor out the overall $M$ dependence of the above integrals. So, in what follows, we will imply $M=1$. After that, the integrals depend only on $w\equiv v_1\cdot v_2$ scalar product.

Similarly, in the case of heavy-light vertex, the required prototype topologies in  Figs.~\ref{fig:top1} $(b)$ can be assembled into the following single integral family 
\begin{equation}
J^{HL}_{\nu_1,\ldots, \nu_{11}} = \int \frac{d^dl_1 d^dl_2}{(i\pi^{d/2})^2}\prod_{i=1}^{11}\frac{1}{(D_i+i0)^{\nu_i}}\, , \label{JHL-prototype}
\end{equation}
where
\begin{align}
& \quad D_1=(l_2+p_2)^2-m^2, \quad D_2=(l_1+p_2)^2-m^2, \quad D_3=l_1\cdot v_1, \quad D_4=(l_1-l_2)\cdot v_1, \nonumber \\&  \quad D_5=l_2^2,   \quad D_6=(l_1-l_2)^2,  \quad D_7=l_2\cdot v_1, \quad D_8=l_1^2, \quad D_9=(l_2+p_2)^2, \nonumber \\&  \quad D_{10}=l_2^2-m^2, \quad D_{11}=(l_1-l_2)^2-m^2.
\end{align}
Here $v_1$ is the heavy field velocity, $p_2$ and $m$ are the full theory field momentum and mass. Again, upon rescaling of integration momenta the overall dependence on $m$ can be factored out and the rescaled integrals depend only on the scalar product $w\equiv v_1\cdot p_2/m$ and we will again imply that in the above integral family definition $m=1$. 

Lastly, the self-energy diagrams which contribute to heavy field renormalization and residual mass term are examined~\cite{manohar2007heavy}. The prototype topologies are shown in Fig.~\ref{fig:top3}, and the corresponding single integral family is defined as 
\begin{equation}
J^{SE}_{\nu_1,\ldots, \nu_{8}} = \int \frac{d^d l_1 d^d l_2}{(i\pi^{d/2})^2}\prod_{i=1}^{8}\frac{1}{(D_i+i0)^{\nu_i}}\, , \label{JSE-prototype}
\end{equation}
where
\begin{align}
& \quad D_1=(p+l_1)\cdot v, \quad D_2=l_2\cdot v, \quad D_3=l_1^2-M^2, \quad D_4=l_2^2-M^2, \nonumber \\& \quad D_5=(l_1-l_2)^2-M^2,  \quad D_6=(p+l_1+l_2)\cdot v,  \quad D_7=l_2^2,  \quad D_8=(l_1-l_2)^2.
\end{align}
Here $p$ and $v$ are the heavy particle residual momentum and velocity and $M$ is the mass of exchanged field. The overall  $M$-dependence can be again factored out, so in what follows we will assume $M=1$. The left integrals depend only on the scalar product $w\equiv v\cdot p/M$. Note, that the relation between the heavy field self-energy, $\Sigma(p)$, the bare field counter-term $\delta Z_h$ and the residual heavy field mass, $\delta m_h$, are given by,
\begin{align}
    \delta Z_{h}={}&i\partial_{v\cdot p}\tilde\Sigma|_{v\cdot p=0} \\
    \delta m_h={}&-i\tilde\Sigma|_{v\cdot p=0}.
\end{align}
To determine these quantities, one only requires the MIs on-shell at $v\cdot p=0$, thus eliminating the momentum, $p$, from the propagators. The resulting MIs are, therefore, simple enough to evaluate with standard techniques. Maintaining $v\cdot p\neq 0$ is interesting in the case of off-shell studies. However, as we will see, in this case, we encounter integrals with elliptic structure.

With the use of IBP relations \cite{IBP1,IBP2} all integrals in the described integral families can be reduced to the set of so called IBP master integrals. To evaluate the latter it is convenient to use the method of differential equations  \cite{diffeqn1,diffeqn2,diffeqn3,diffeqn4,diffeqn5}. As all master integrals we consider are dependent on a particular scalar product $w$, it is natural to consider differential equations for our master integrals with respect to $w$. To take the derivative with respect to an arbitrary scalar product $ p \cdot q$, for two arbitrary vectors $p$ and $q$, one can follow one of two equivalent ways,
\begin{equation}
    \frac{\partial}{\partial(p\cdot q)}=\frac{(p\cdot q) p-p^2q}{(p\cdot q)^2-p^2q^2}\cdot\frac{\partial}{\partial p}=\frac{(p\cdot q) q-q^2p}{(p\cdot q)^2-p^2q^2}\cdot\frac{\partial}{\partial q}
\end{equation}
In our study, we take derivatives with respect to the parameter $w$, defined in each integral family considered. Upon re-reducing the differentiated results with IBP identities, we obtain a linear combination of MIs, leading to a set of coupled differential equations. More precisely, the derivative of a given MI will inevitably lie in the same sector or sub-sector, meaning they contain the same set of non-zero $\nu_i$, or a smaller set, compared to the original MI. Thus, one can combine all MIs and their derivatives into a linear system of differential equations.
As we will see below, these systems can be further reduced either to $\ep$ or $A+B\ep$ forms in the cases of vertex and self-energy integral families, respectively.   From there, we solve each system iteratively at each order in a Laurent expansion about small $\ep$.


\subsection{Heavy-heavy vertex}
\begin{figure}[tb]
	\begin{center}
		\includegraphics[width=15cm]{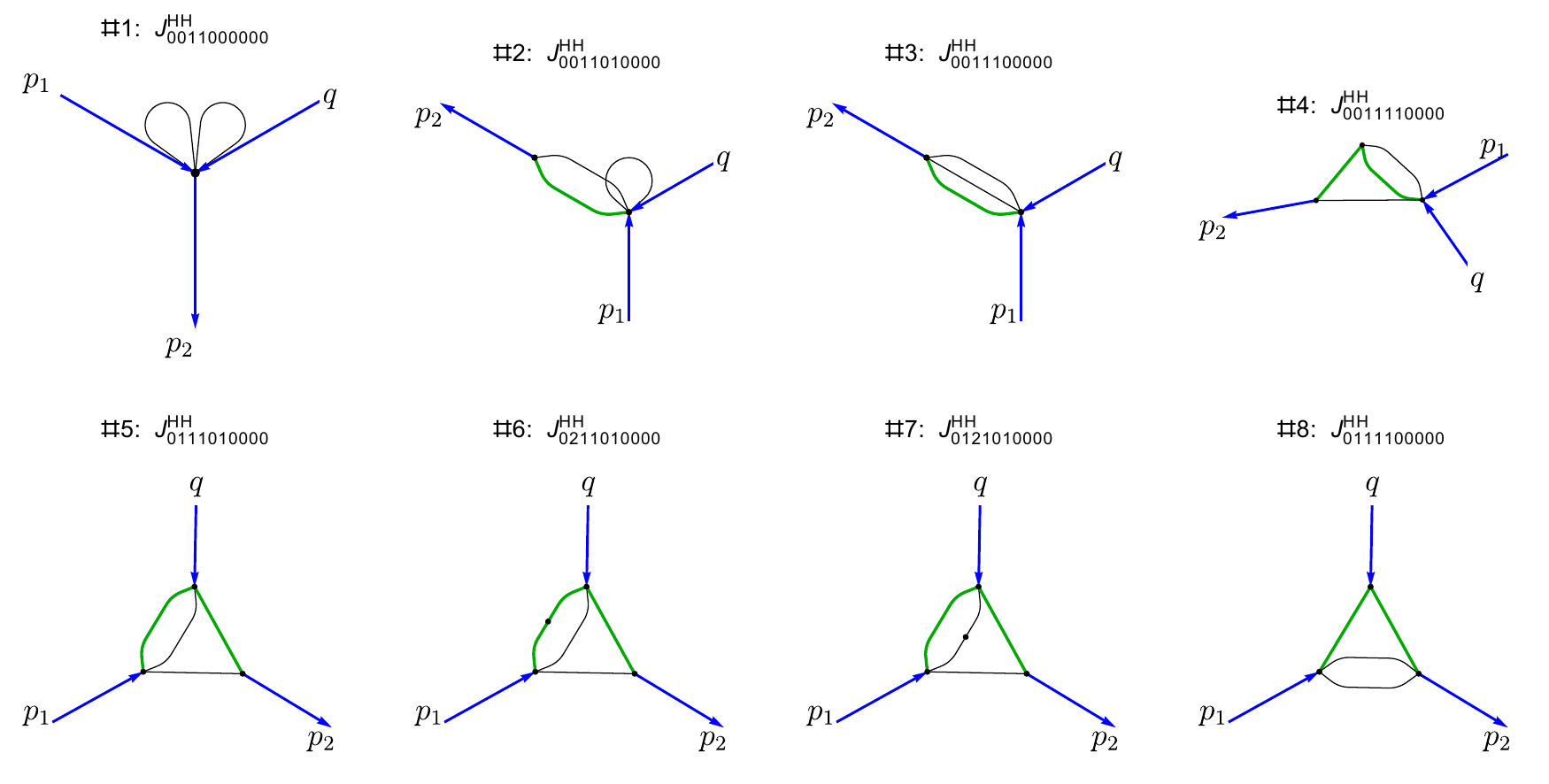}		
		\caption{Master integrals for heavy-heavy vertex (\#: 1-8). Green lines denote propagators for heavy particles, solid - propagators for massive particles and dashed - massless propagators.}
		\label{fig:mastersHHa}
	\end{center}
\end{figure}
\begin{figure}[tb]
	\begin{center}
		\includegraphics[width=15cm]{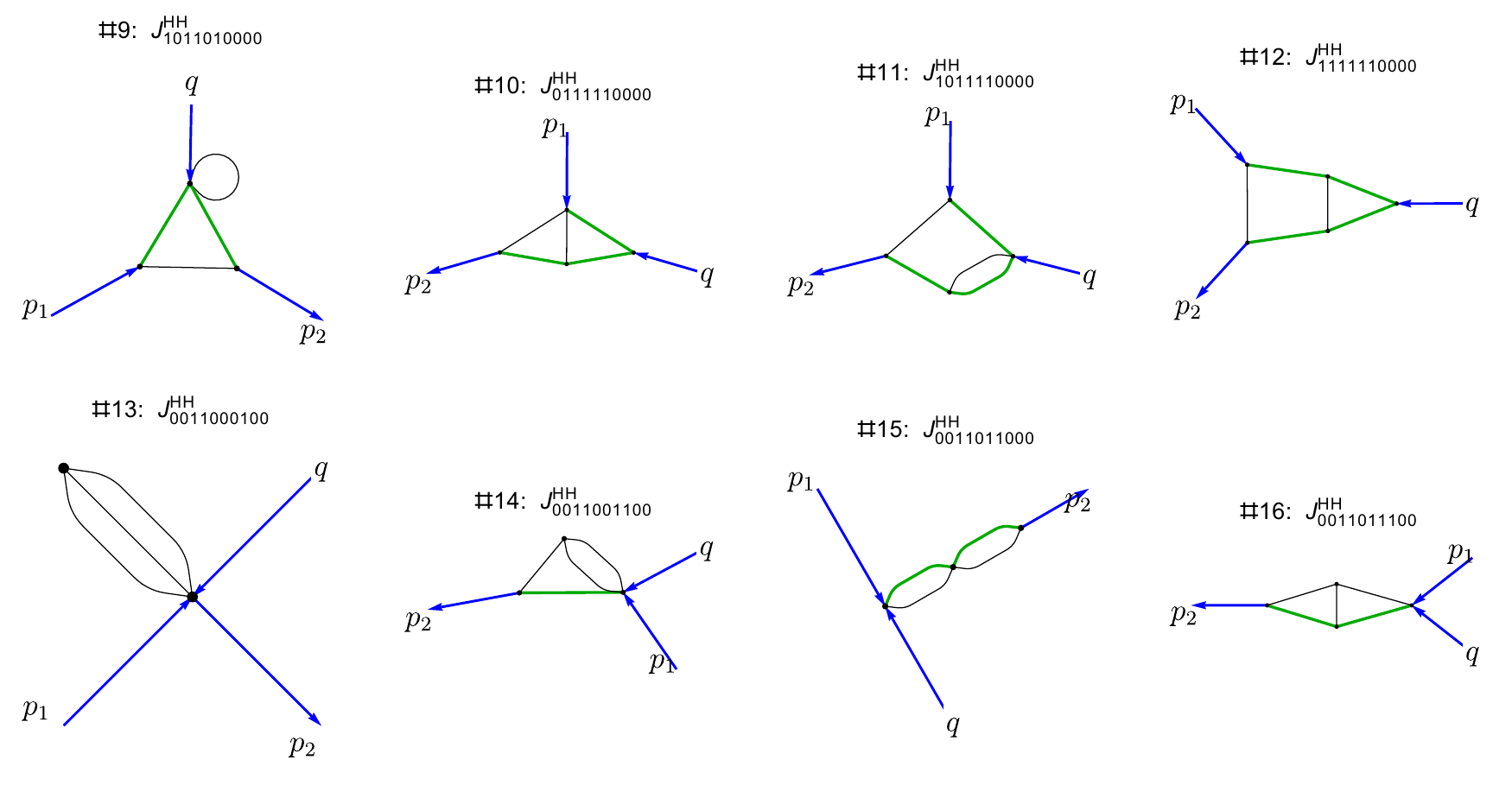}		
		\caption{Master integrals for heavy-heavy vertex (\#: 9-16). Green lines denote propagators for heavy particles, solid - propagators for massive particles and dashed - massless propagators.}
		\label{fig:mastersHHb}
	\end{center}
\end{figure}
\begin{figure}[tb]
	\begin{center}
		\includegraphics[width=15cm]{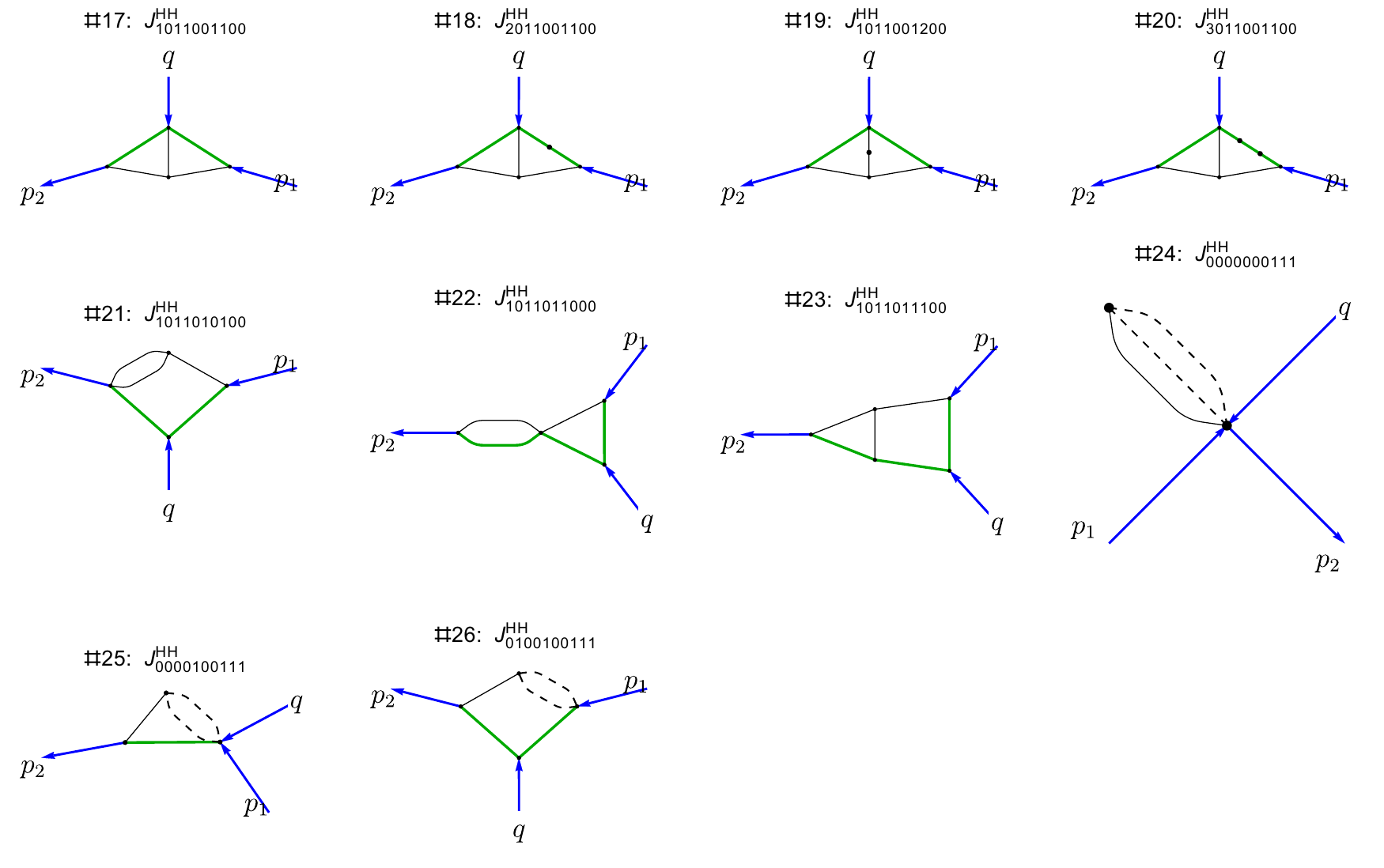}		
		\caption{Master integrals for heavy-heavy vertex (\#: 17-26). Green lines denote propagators for heavy particles, solid - propagators for massive particles and dashed - massless propagators.}
		\label{fig:mastersHHc}
	\end{center}
\end{figure}

In the heavy-heavy vertex case we have 26 master integrals shown in Figs.~\ref{fig:mastersHHa}-\ref{fig:mastersHHc}. The latter could be conveniently represented as a column vector  $\boldsymbol{A}(w)$ with component $A_i(w)$ corresponding to $J_{\nu_1,\ldots ,\nu_{10}}^{HH}(w)$ master $\# i$. Upon differentiation of $\boldsymbol{A}(w)$ with respect to $w$ and reduction by IBP identities, we have a differential system,
\begin{equation}
\partial_w\boldsymbol{A}(w)=\mathbb{M}(w,\ep)\boldsymbol{A}(w).
\end{equation}
with a $26\times 26$ matrix, $\mathbb{M}(w,\ep)$ which is neither Fuchsian nor in $\ep$-form. Using balance transformations and the algorithm outlined in \cite{epform2} this system can be however transformed into $\ep$-form. First, we transform to $\ep$-form diagonal blocks. The largest diagonal block (for masters $\#17,\#19$ and $\#20$) is given by $3\times 3$ matrix 
\begin{equation}
    M_3=\left(
\begin{array}{ccc}
 -\frac{2 w^2-\ep }{2 (w-1) w (w+1)} & \frac{3}{4 (w-1) w (w+1)} & \frac{(2 w-1) (2 w+1)}{4 (w-1) w (w+1) \ep } \\
 \frac{\ep ^2}{(w-1) w (w+1)} & -\frac{2 w^2-3 \ep }{2 (w-1) w (w+1)} & -\frac{1}{2 (w-1) w (w+1)} \\
 \frac{2 \ep ^2 \left(2 w^2 \ep -2 w^2-2 \ep -1\right)}{(w-1) w (w+1) (2 w-1) (2 w+1)} & \frac{3 \ep  \left(2
   w^2 \ep +2 w^2-2 \ep -1\right)}{(w-1) w (w+1) (2 w-1) (2 w+1)} & -\frac{4 w^4-4 w^2 \ep -3 w^2-2 \ep
   -1}{(w-1) w (w+1) (2 w-1) (2 w+1)} \\
\end{array}
\right).
\end{equation}
Using  \texttt{Libra} package \cite{Libra} we can easily find the sequence of balance transformations to transform eigenvalues of $M_3$ matrix residues at all poles except at $w=\pm 1$ to $n\ep$ form, where $n$ is some integer. The required overall transformation matrix is given by
\begin{equation}
   T_3= \left(
\begin{array}{ccc}
 \frac{1}{w+1} & 0 & 0 \\
 0 & \frac{1}{w-1} & 0 \\
 \frac{2 \ep ^2}{w+1} & -\frac{3 \ep }{(w-1)(4w^2-1)} & \frac{4 w}{4 w^2-1} \\
\end{array}
\right),
\end{equation}.
The eigenvalues at $w=\pm 1$ have the form $\pm 1/2+n\ep$ and to normalize them further we require variable change. It is clear\footnote{See also \cite{epform-criterium}.} that the appropriate variable employ is 
\begin{equation}
w = \frac{1-\beta^2}{1+\beta^2}\, , \quad \beta = \sqrt{\frac{1-w}{1+w}} \label{beta-def}
\end{equation}
Now, we again use  \texttt{Libra} to find the required sequence of balance transformations to normalize eigenvalues at $\beta=0,\infty$ to $n\ep$ form and find $\beta$-independent transformation to factor out the overall $\ep$-dependence. The required transformation matrix was found to be 
\begin{equation}
    T_3'=\left(
\begin{array}{ccc}
 \frac{1}{\beta } & 0 & 0 \\
 0 & \frac{\beta  \ep }{2} & 0 \\
 0 & 0 & \frac{\ep ^2}{4} \\
\end{array}
\right),
\end{equation}
which together with previously found matrix $T_3$ reduces $M_3$ to $\ep$-form,
\begin{equation}
S_3  = \ep \left(
\begin{array}{ccc}
\frac{2w}{w^2-1} & 0   & \frac{\beta}{4(w-1)} \\
0 & \frac{6w}{(w^2-1)(4w^2-1)} & \frac{\beta}{(w-1)(4w^2-1)}  \\
\frac{16\beta}{w-1} & \frac{12\beta}{(w-1)(4w^2-1)}  & - \frac{8w}{4w^2-1}  \\
\end{array}
\right).
\end{equation}
Next, we repeat these steps for other diagonal blocks. Finally, fuchsifying  off-diagonal blocks and factoring out overall $\ep$-dependence\footnote{See \cite{epform2} for detail description of these steps.}  the original system of differential equations is reduced to canonical or  $\ep$-form:
\begin{equation}
\partial_w\tilde{\boldsymbol{A}}(w)=\ep\mathbb{S}(w)\tilde{\boldsymbol{A}}(w)\, , 
\end{equation}
where $\boldsymbol{A}=\mathbb{T}\cdot\tilde{\boldsymbol{A}}$. In what follows we will refer to the vector of master integrals $\tilde{\boldsymbol{A}}$ as the vector of canonical master integrals. It turns out that in this case the only new variable appearing in the process of reduction is $\beta$ and thus it is more convenient to consider differential system with respect to $\beta$
\begin{equation}
\partial_\beta\tilde{\boldsymbol{A}}(\beta)=\ep\tilde{\mathbb{M}}(\beta)\tilde{\boldsymbol{A}}(\beta)\, , 
\end{equation}
This way, as we will see in the next section, the corresponding solution of the differential system can be written in terms of MPLs. The corresponding expressions for the canonical $\tilde{\mathbb{M}}$
and transformation matrices $\mathbb{T}$ can be found in the accompanying {\it Mathematica} notebook.


\subsection{Heavy-light vertex}
\begin{figure}[tb]
	\begin{center}
		\includegraphics[width=15cm]{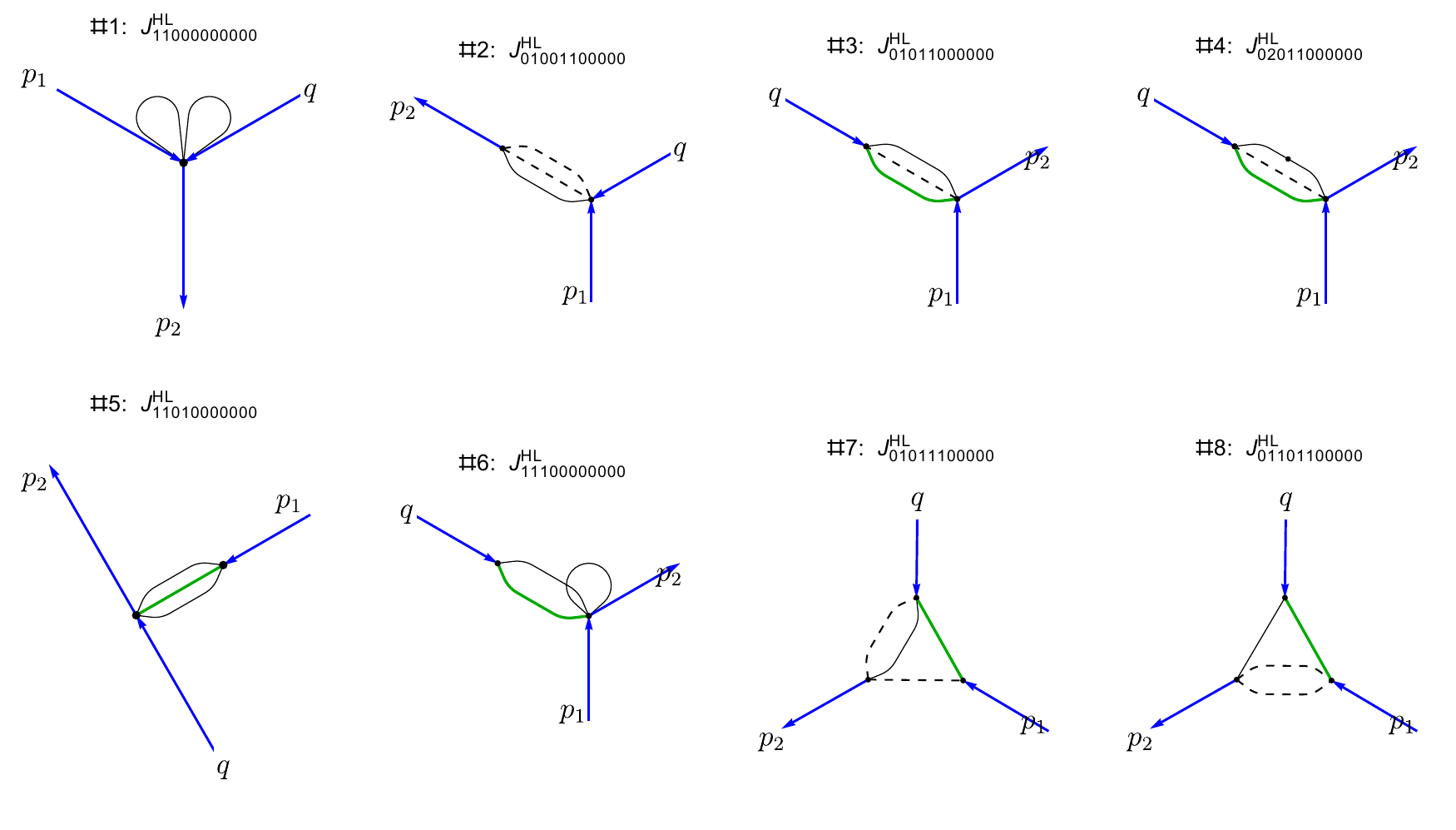}		
		\caption{Master integrals for heavy-light vertex (\#: 1-8). Green lines denote propagators for heavy particles, solid - propagators for massive particles and dashed - massless propagators.}
		\label{fig:mastersHLa}
	\end{center}
\end{figure}
\begin{figure}[tb]
	\begin{center}
		\includegraphics[width=15cm]{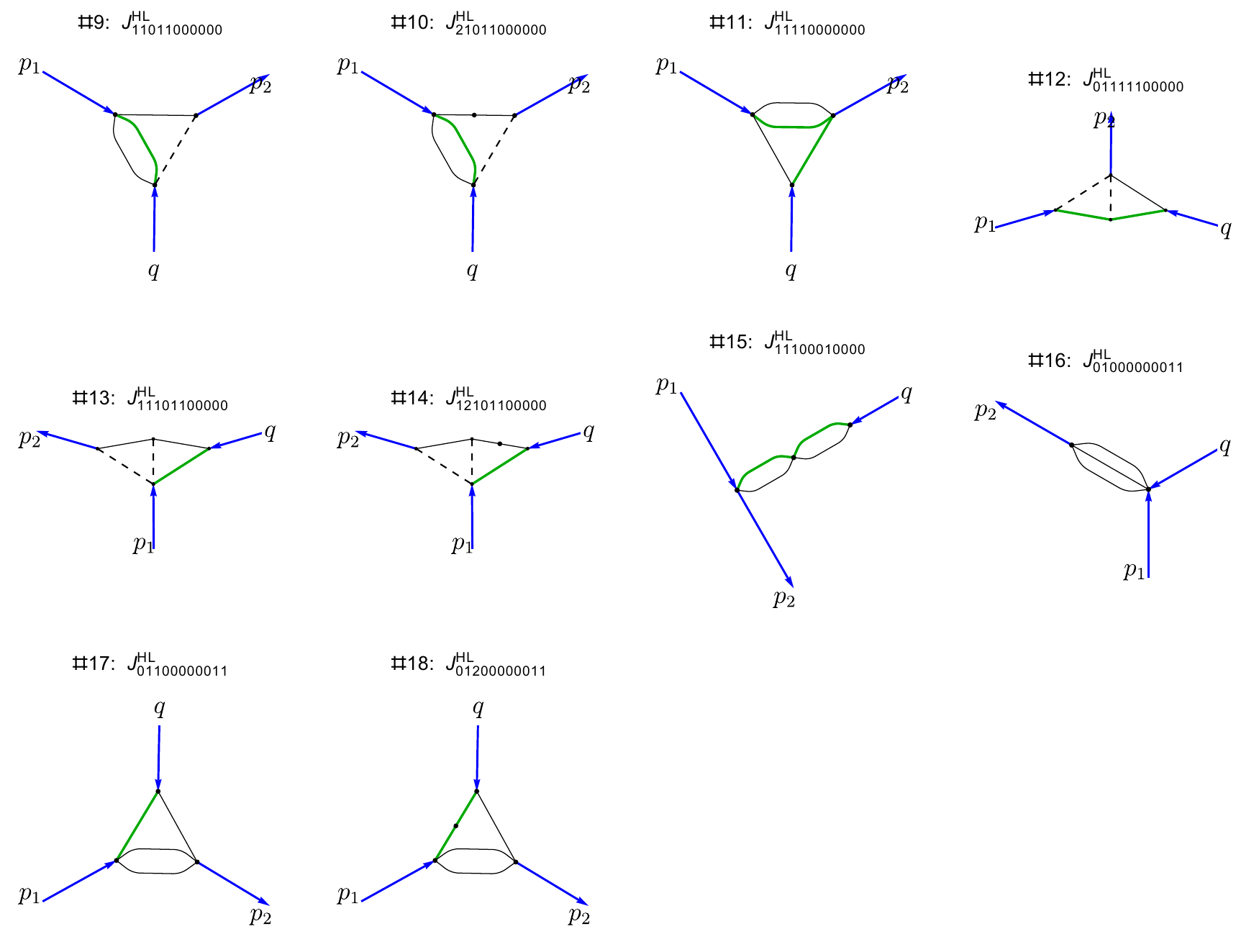}		
		\caption{Master integrals for heavy-light vertex (\#: 9-18). Green lines denote propagators for heavy particles, solid - propagators for massive particles and dashed - massless propagators.}
		\label{fig:mastersHLb}
	\end{center}
\end{figure}

Here, we have 18 master integrals shown in Figs.~\ref{fig:mastersHLa} and \ref{fig:mastersHLb}. Writing the latter as a column vector $\boldsymbol{B}(w)$ the corresponding differential equations system can be written as 
\begin{equation}
\partial_w\boldsymbol{B}(w)=\mathbb{M}(w,\ep)\boldsymbol{B}(w).
\end{equation}
with a $18\times 18$ matrix, $\mathbb{M}(w,\ep)$. The reduction to $\ep$-form in the present case proceeds similar to the case of heavy-heavy vertex. That is, we first reduce to $\ep$-form diagonal blocks. Then, we fuchsify off-diagonal blocks and finally factor out the overall $\ep$-dependence. What is important is that in this case similar to the case of heavy-heavy vertex the only new variable introduced is $\beta$ \eqref{beta-def}. For example, take $2\times 2$ diagonal block (for masters $\# 13$ and $\# 14$)
\begin{equation}
    M_2=\left(
\begin{array}{cc}
 -\frac{2 w^2 \ep +w^2-4 \ep }{(w-1) w (w+1)} & -\frac{4}{w} \\
 -\frac{\ep  (4 \ep +1)}{(w-1) w (w+1)} & -\frac{w^2+4 \ep +1}{(w-1) w (w+1)} \\
\end{array}
\right).
\end{equation}
Using balance transformations we again transform eigenvalues of $M_2$ matrix residues at all poles except at $w = \pm 1$ to $n\ep$ form, where $n$ is some integer. This can be  done for example with transformation matrix 
\begin{equation}
T_2=\left(
\begin{array}{cc}
\frac{1}{w+1} & 0 \\
-\frac{(w-2) \ep }{2 \left(w^2-1\right)} & \frac{w}{w^2-1} \\
\end{array}
\right).
\end{equation}
To reduce eigenvalues at $w=\pm 1$ we require the same variable change to $\beta$-variable \eqref{beta-def} as in the case of heavy-heavy vertex.  Next, we again use  \texttt{Libra} to find a sequence of balance transformations  to normalize eigenvalues at $\beta=0,\infty$ to $n\ep$ form and find $\beta$-independent transformation to factor out the overall $\ep$-dependence. The required transformation matrix was found to be 
\begin{equation}
T_2'=\left(
\begin{array}{cc}
\frac{1}{\beta } & 0 \\
-\frac{\ep}{2 \beta } & \frac{\ep}{2} \\
\end{array}
\right),
\end{equation}
which together with previously found matrix $T_2$ transforms $M_2$ to $\ep$-form,
\begin{equation}
S_2  = \ep \left(
\begin{array}{cc}
\frac{2w}{w^2-1} & -\frac{2\beta}{w-1}   \\
-\frac{4\beta}{w-1} & -\frac{4w}{w^2-1}   \\
\end{array}
\right).
\end{equation}
Altogether, in this case we can also write down the transformed differential system with respect to $\beta$ in $\ep$-form ($\boldsymbol{B}=\mathbb{T}\cdot\tilde{\boldsymbol{B}}$)
\begin{equation}
\partial_\beta\tilde{\boldsymbol{B}}(\beta)=\ep\tilde{\mathbb{M}}(\beta)\tilde{\boldsymbol{B}}(\beta)\, ,
\end{equation}
with the corresponding expressions for canonical $\tilde{\mathbb{M}}$
and transformation $\mathbb{T}$ matrices located in the accompanying {\it Mathematica} notebook.

\subsection{Heavy propagator}
\begin{figure}[tb]
	\begin{center}
		\includegraphics[width=15cm]{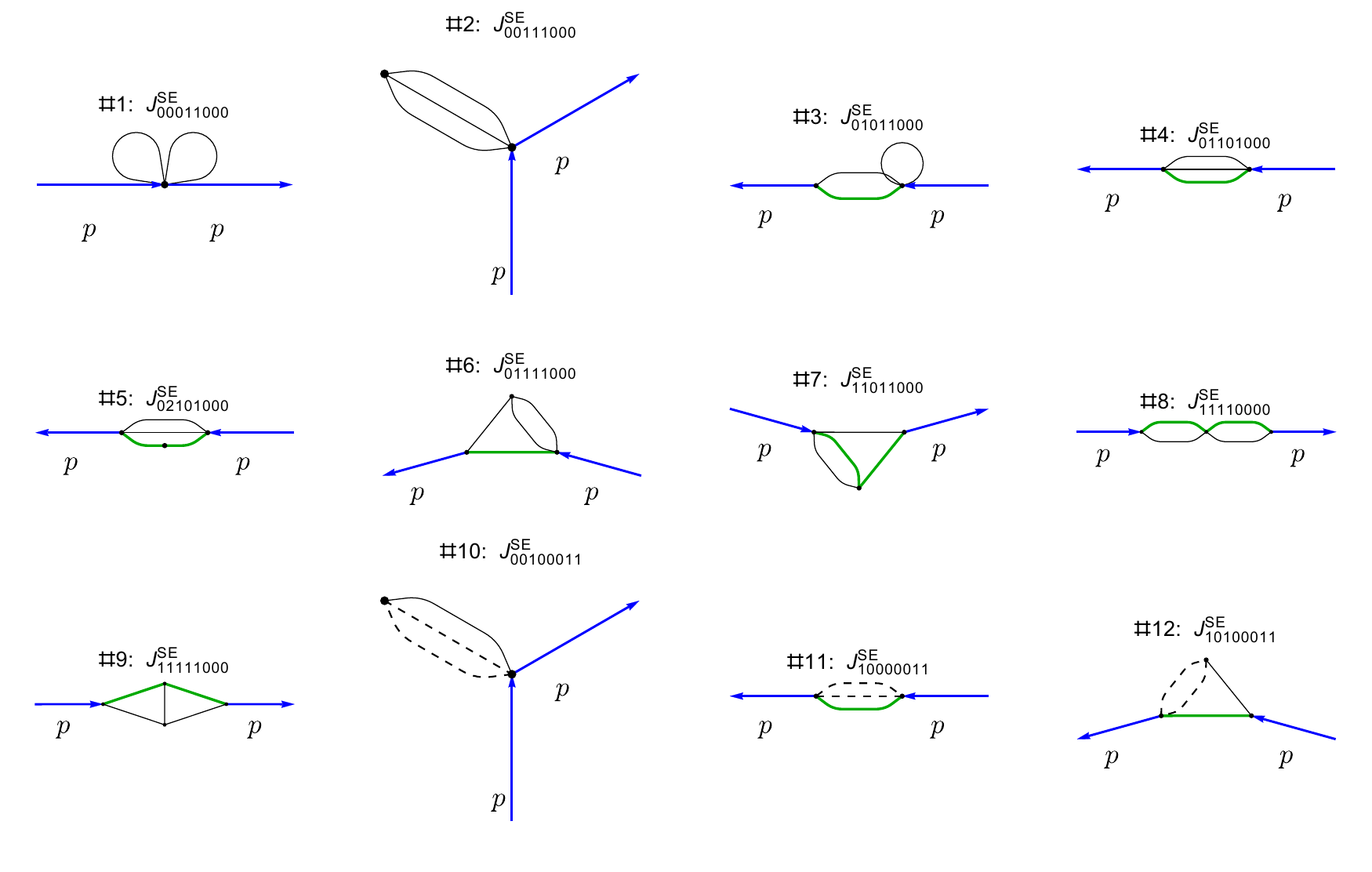}		
		\caption{Master integrals for heavy propagator (\#: 1-12). Green lines denote propagators for heavy particles, solid - propagators for massive particles and dashed - massless propagators.}
		\label{fig:mastersSE}
	\end{center}
\end{figure}

In the case of the heavy propagator, we have 12 master integrals as depicted in Fig.~\ref{fig:mastersSE}. The reduction of the corresponding differential system for a vector of master integrals $\boldsymbol{C}$ is achieved as in the cases of heavy-heavy and heavy-light vertices. In particular, we first reduce diagonal blocks where possible to $\ep$-form and then fuchsify off-diagonal blocks. Some of the diagonal blocks will also require a variable change to the $\beta$ variable, as defined in \eqref{beta-def}. However, contrary to the vertex cases considered previously, we have elliptic sub-graphs represented by elliptic sunsets (master integrals $\# 4$ and $\# 5$). This particular diagonal block can not be reduced to $\ep$-form\footnote{See \cite{epform-criterium} for the criterion of such reducibility.}. Still, one can reduce the latter to $A+B\ep$ form. Also, instead of factoring out the overall $\ep$-dependence in the final step, we find a ($\beta (w)$-independent) diagonal transformation matrix, which reduces the entire system to $A+B\ep$ form. Altogether, in this case we write down the transformed differential system with respect to $w$ in $A+B\ep$ form ($\boldsymbol{C}=\mathbb{T}\cdot\tilde{\boldsymbol{C}}$): 
\begin{equation}
\partial_w\tilde{\boldsymbol{C}}(w)=\tilde{\mathbb{M}}(w)\tilde{\boldsymbol{C}}(w)\, ,
\end{equation}
with corresponding expressions for canonical $\tilde{\mathbb{M}}$
and transformation $\mathbb{T}$ matrices located in accompanying {\it Mathematica} notebook. Here, the matrix $\tilde{\mathbb{M}}$ does contain square roots in $w$ induced by the variable $\beta$. As we already noted, for on-shell studies we need to know master integrals only for $w=0$. Still, as will be shown in next section, we will  also able to supply all the ingredients necessary to have Frobenius solution in terms of a generalized power series in $w$ for $w\neq 0$.   

\begin{figure}[b]
	\begin{center}
		\includegraphics[width=15cm]{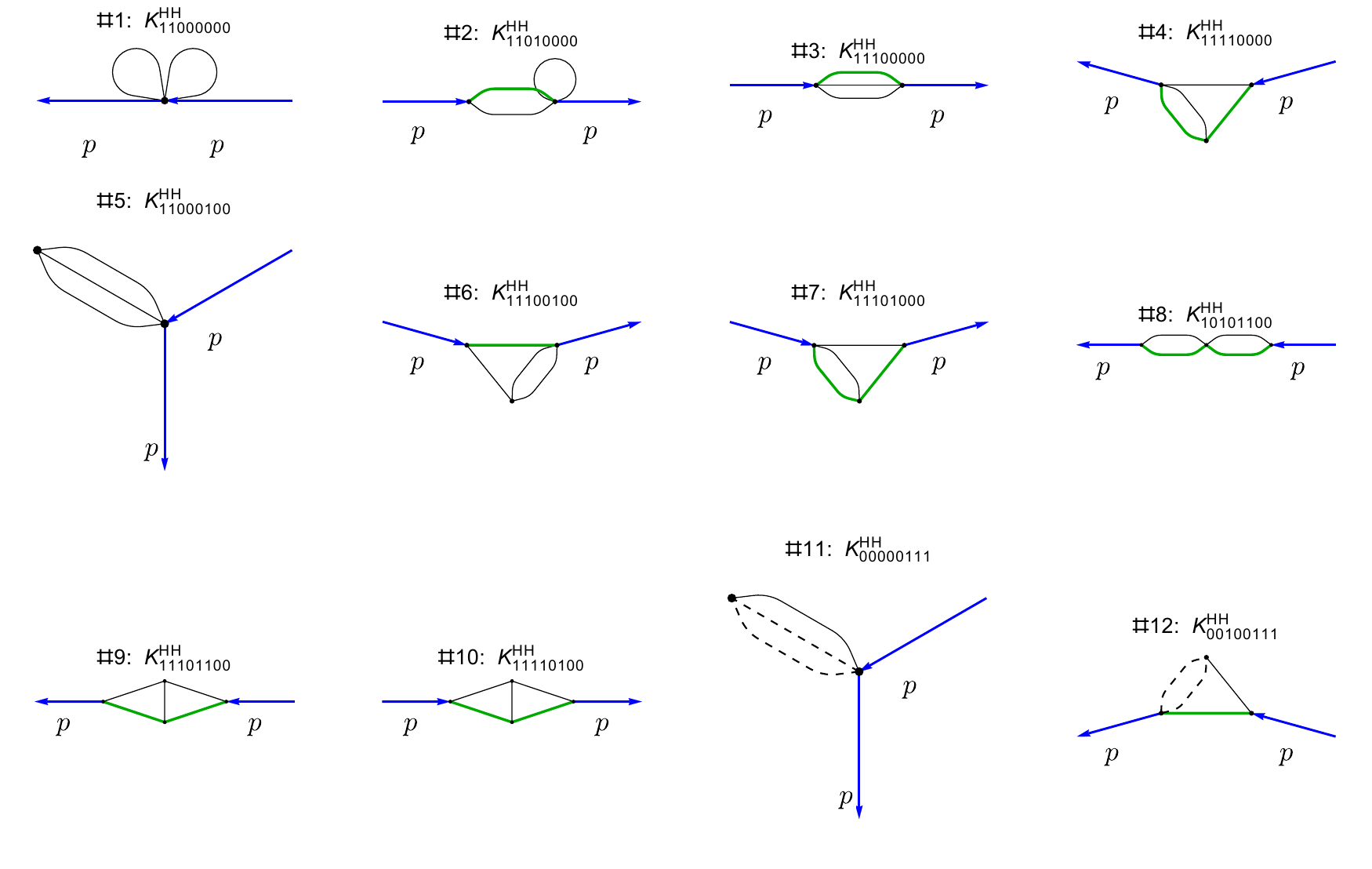}		
		\caption{Boundary master integrals for heavy-heavy vertex (\#: 1-12). Green lines denote propagators for heavy particles, solid - propagators for massive particles and dashed - massless propagators.}
		\label{fig:mastersHH-boundary}
	\end{center}
\end{figure}

\section{Solution of differential equations and results}
\label{scn:results}
\begin{figure}[b]
	\begin{center}
		\includegraphics[width=15cm]{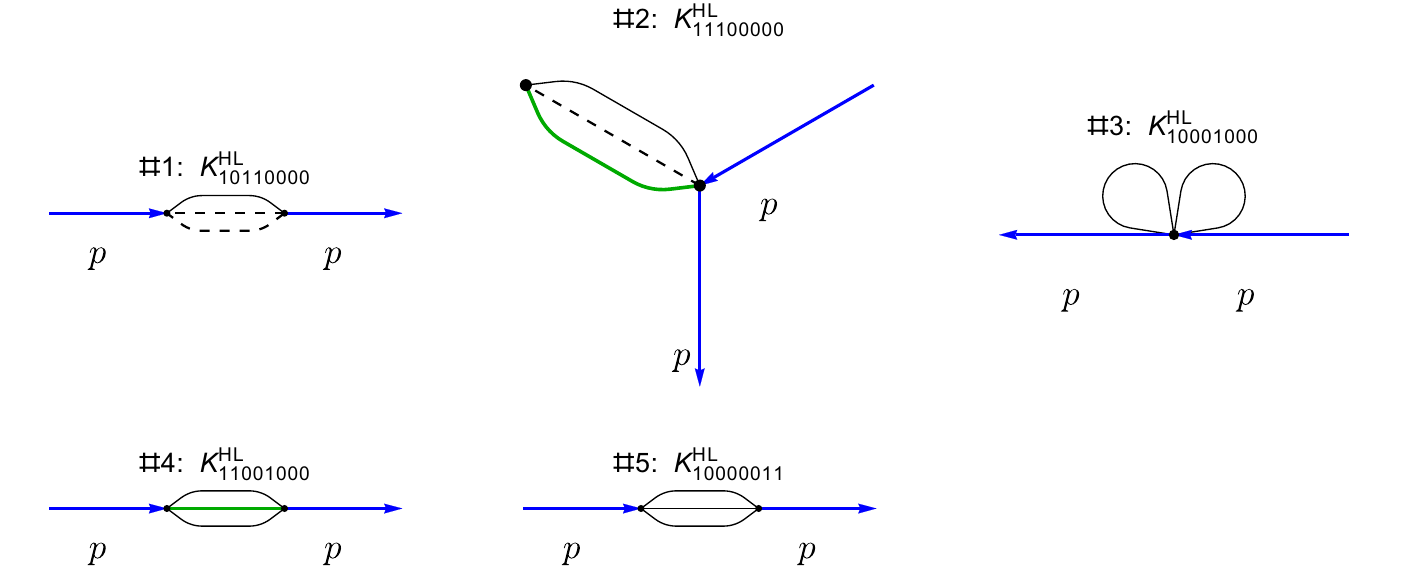}		
		\caption{Boundary master integrals for heavy-light vertex (\#: 1-5). Green lines denote propagators for heavy particles, solid - propagators for massive particles and dashed - massless propagators.}
		\label{fig:mastersHL-boundary}
	\end{center}
\end{figure}
\begin{figure}[b]
	\begin{center}
		\includegraphics[width=15cm]{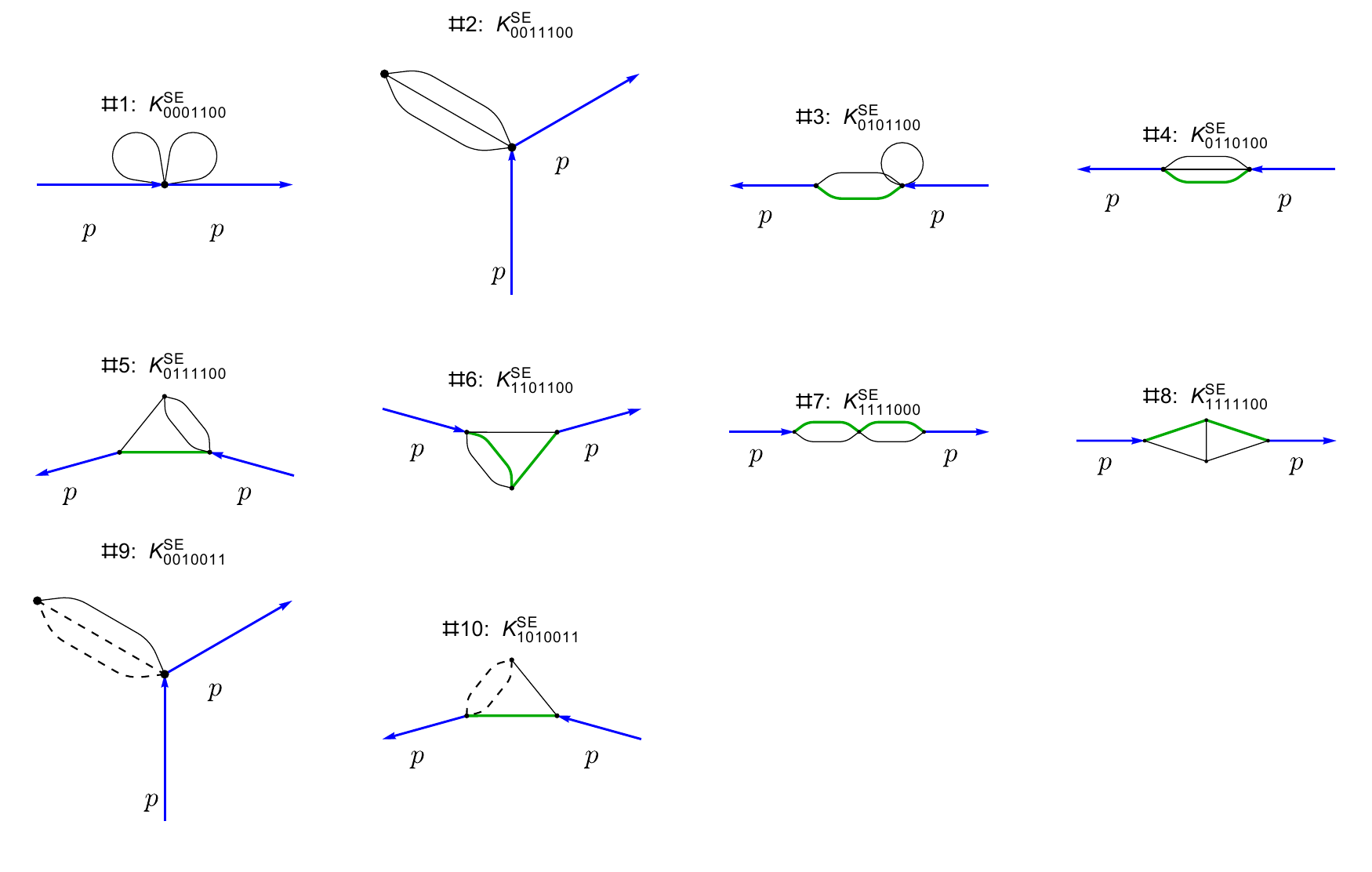}		
		\caption{Boundary master integrals for heavy propagator (\#: 1-10). Green lines denote propagators for heavy particles, solid - propagators for massive particles and dashed - massless propagators.}
		\label{fig:mastersSE-boundary}
	\end{center}
\end{figure}

Given our reduced systems of differential equations to either $\ep$ or $A+B\ep$ form in the previous section, we are now ready to solve them. In the case of heavy-heavy and heavy-light vertices, where we managed to reduce the corresponding differential systems to $\ep$-form, the solution can be readily written in terms of $\mathcal{P}$-exponents as
\begin{equation}
\boldsymbol{J}(\beta)=\mathbb{T}(\beta)  \tilde{\boldsymbol{J}}(\beta)=\mathbb{T}(\beta)\mathcal{P}\exp{\left[\ep\int_0^\beta\tilde{\mathbb{M}}(t)dt\right]\mathbb{L}\cdot \boldsymbol{c}}\, ,
\end{equation}
where $\boldsymbol{J}$ is either vector of $\boldsymbol{A}$ or $\boldsymbol{B}$ masters, $\mathbb{T}$ is the transformation matrix to canonical basis and $\tilde{\mathbb{M}}$ is corresponding differential equations matrix in canonical basis. $\mathbb{L}$ is the adapter matrix converting the vector of constants $\boldsymbol{c}$ into boundary conditions for canonical master integrals. The expressions for said boundary conditions can be found in the accompanying {\it Mathematica} notebook. The components of the vector of constants $\boldsymbol{c}$ have the form $c_i(\beta^j)$, where $c_i(\beta^j)$ is the coefficient of $\beta^j$ in the $\beta$-expansion of the $i$-th master. In the case of the heavy-heavy and heavy-light vertices, $\boldsymbol{c}$-vectors are given by
\begin{multline}
\boldsymbol{c}_{HH} = \Big(
c_1(\beta^0), c_2(\beta^0), c_3(\beta^0), c_6 (\beta^{-1+2\ep}), c_5(\beta^0), c_5(\beta^{-1+2\ep}), c_8(\beta^{-1}), c_9(\beta^{-1}), c_{10}(\beta^{-1}), \\
c_{11}(\beta^{-1}), c_{12}(\beta^{-2}), c_{13}(\beta^0), c_{14}(\beta^0), c_{15}(\beta^0),
c_{16}(\beta^0), c_{18}(\beta^{-1+2\ep}), c_{17}(\beta^0), c_{17}(\beta^{-1+2\ep}), \\
c_{17}(\beta^{1+2\ep}), c_{21}(\beta^{-1}), c_{22}(\beta^{-1}), c_{23}(\beta^{-1}), c_{24}(\beta^0), c_{4}(\beta^0), c_{25}(\beta^0), c_{26}(\beta^{-1})
\Big)^\top
\end{multline} 
and 
\begin{multline}
\boldsymbol{c}_{HL} = \Big(
c_1(\beta^0), c_2(\beta^0), c_3(\beta^0), c_3(\beta^{5-6\ep}), c_5(\beta^0), c_6(\beta^{1-2\ep}), c_7(\beta^{-1+2\ep}), c_8(\beta^{1-4\ep}), \\ 
c_9(\beta^{-1+2\ep}), c_{10}(\beta^{-4\ep}), c_{11}(\beta^{2-4\ep}), c_{12}(\beta^{-4\ep}),
c_{13}(\beta^{-4\ep}), c_{13}(\beta^{-1+2\ep}), \\ 
c_{15}(\beta^{2-4\ep}), c_{16}(\beta^0), c_{17}(\beta^{-2\ep}), c_{17}(\beta^{-1+2\ep})
\Big)^\top\, .
\end{multline} 
It is easy to see, that all  $c_i(\beta^j)$ constants other than $c_i(\beta^0)$ are zero either due to the regularity of master integrals at $\beta = 0$ or due to the absence of sub-graphs in the large mass expansion (corresponding to the expansion at $\beta = 0$) of these vertices. The remaining constants will be calculated in the next sub-section with the use of dimensional recursion relations.

In the heavy propagator case, due to the presence of elliptic sub-graphs in its corresponding differential system, a $\mathcal{P}$-exponent solution is not attainable. However, we can still obtain a Frobenius solution in terms of a generalised power series in the  $w$-variable. The solution can be written as
\begin{equation}
\boldsymbol{C}(w)=\mathbb{T}(w)  \tilde{\boldsymbol{C}}(w)=\mathbb{T}(w)\mathcal{F}(w)\cdot \mathbb{L}\cdot \boldsymbol{c} ,
\end{equation}
where $\mathcal{F}(w)$ is the Frobenius solution for canonical master integrals. $\mathbb{L}$ is again the adapter matrix converting the vector of constants $\boldsymbol{c}$ into boundary conditions for canonical master integrals. the vector of constants $\boldsymbol{c}$ in this case is given by
\begin{multline}
\boldsymbol{c}_{SE} = \Big(
c_{1}(\beta^0), c_{2}(\beta^0), c_3(\beta^0), c_4(\beta^0), c_4(\beta^{2-2\ep}), c_6(\beta^0), c_7(\beta^0), c_8(\beta^0), c_9(\beta^0), \\
 c_{10}(\beta^0), c_{11} (\beta^{3-4\ep}), c_{12}(\beta^0)
\Big)^\top\, .
\end{multline} 
Using Feynman parameters it is easy to see that $c_4 (\beta^{2-2\ep}) = 0$ and 
\begin{equation}
c_{11}(\beta^{3-4\ep}) = (-1)^{4\ep} 2^{4-4\ep} \Gamma (1-\ep)^2\Gamma (4\ep -3)\, .
\end{equation}
Like the case of vertices, the remaining constants are more convenient to determine using dimensional recurrence relations. It is also possible to determine Frobenius solutions for canonical master integrals $\mathcal{F}(w)$, see for example \cite{frobenius1,frobenius2,frobenius3,frobenius4,frobenius5,frobenius6,frobenius7}. Moreover, the \texttt{Libra} package \cite{Libra} itself contains built-in tools for obtaining such a solution. The reader is advised to consult the accompanying {\it Mathematica} notebook.

\subsection{Boundary constants from dimensional recurrences}

We need to calculate the remaining $c_i(\beta^0)$ constants in each case. First, we note that for $\beta = 0$ in the case of vertices, or $w=0$ in the case of heavy propagators, the master integrals, as illustrated in Figs.~\ref{fig:mastersHHa}-\ref{fig:mastersSE}, are no longer masters and additional reduction using partial fractioning and IBP identities is possible. In the case of the heavy-heavy vertex, additional boundary master integrals can be written in terms of a single integral family
\begin{equation}
K^{HH}_{\nu_1,\ldots, \nu_{8}} = \int \frac{d^d l_1 d^d l_2}{(i\pi^{d/2})^2}\prod_{i=1}^{8}\frac{1}{(D_i+i0)^{\nu_i}}\, , \label{KHH-prototype}
\end{equation}
where
\begin{align}
& \quad D_1=(l_1-l_2)^2-1, \quad D_2=l_2^2-1, \quad D_3=l_1\cdot v,\quad  D_4=l_2\cdot v,   \nonumber \\& \quad D_5=(l_2-l_1)\cdot v, \quad D_6=l_1^2-1, \quad D_7=l_2^2, \quad D_{8}=(l_1-l_2)^2
\end{align}
and $v$ is the heavy field velocity. Altogether in this case we have 12 master integrals shown in Fig.~\ref{fig:mastersHH-boundary}. Similarly, in the case of heavy-light vertex the single integral family is given by  
\begin{equation}
K^{HL}_{\nu_1,\ldots, \nu_{8}} = \int \frac{d^d l_1 d^d l_2}{(i\pi^{d/2})^2}\prod_{i=1}^{8}\frac{1}{(D_i+i0)^{\nu_i}}\, , \label{KHL-prototype}
\end{equation}
where
\begin{align}
& \quad D_1=(l_1+v)^2-1, \quad D_2=(l_1-l_2)\cdot v, \quad D_3=l_2^2,\quad  D_4=(l_1-l_2)^2,   \nonumber \\& \quad D_5=(l_2+v)^2-1, \quad D_6=l_1^2, \quad D_7=l_2^2-1, \quad D_{8}=(l_1-l_2)^2-1
\end{align}
and the corresponding 5 boundary master integrals are shown in Fig.~\ref{fig:mastersHL-boundary}.
Finally, for the heavy propagator case the single integral family takes the form
\begin{equation}
K^{SE}_{\nu_1,\ldots, \nu_{7}} = \int \frac{d^d l_1 d^d l_2}{(i\pi^{d/2})^2}\prod_{i=1}^{7}\frac{1}{(D_i+i0)^{\nu_i}}\, , \label{SE-prototype}
\end{equation}
where
\begin{align}
& \quad D_1=v\cdot l_1, \quad D_2=v\cdot l_2, \quad D_3=l_1^2-1,\quad  D_4=l_2^2-1,   \nonumber \\& \quad D_5=(l_1-l_2)^2-1, \quad D_6=l_2^2, \quad D_7=(l_1-l_2)^2
\end{align}
with 10 boundary master integrals shown in Fig.~\ref{fig:mastersSE-boundary}.

The simplest way to calculate the exact expressions in $\ep$ for these boundary master integrals is to use dimensional recurrence relations \cite{tarasov1996connection,dimrecSunrise,lee2010space,lee2010calculating}. In all cases we present here, the recurrence relations are all first-order. Consider, for instance, the evaluation of the $K^{SE}_{0011100}(d)$ master integral. In this case, we have the following dimensional recurrence relation ($\nu = d/2$)
\begin{equation}
K^{SE}_{0011100}(\nu+1) = \frac{3}{2\nu (2\nu-1)}K^{SE}_{0011100}(\nu) + \frac{3}{2\nu (2\nu -1)}K^{SE}_{0001100}(\nu)\, ,
\end{equation} 
where $K^{SE}_{0001100}(\nu)$ is the product of two one-loop tadpoles
\begin{equation}
K^{SE}_{0001100}(\nu) = \Gamma (1-\nu)^2\, .
\end{equation}
It is sufficient to consider the solution of this recurrence relation in a basic strip $\nu\in [1,2)$. First, we determine the solution of the homogeneous difference equation,
\begin{equation}
J_h (\nu) = \frac{2\cdot 3^{\nu-1}\pi \csc (2\pi\nu)}{\Gamma (2\nu -1)}\, ,
\end{equation}
where an extra periodic factor $\csc(2\pi\nu)$ was chosen to be well-behaved at imaginary infinities and singularities close to those of the original master integral $K^{SE}_{0011100}(\nu)$ (see \cite{lee2010space,lee2010calculating} for further details). The easiest way to see them is through numerical evaluation in a chosen strip with sector decomposition \cite{Binoth:2000ps,Binoth:2003ak,Binoth:2004jv,Heinrich:2008si,Bogner:2007cr,Bogner:2008ry,Kaneko:2009qx}.  In particular, we used the function \texttt{SDAnalyse} from \texttt{Fiesta} \cite{Fiesta4}. Knowing the homogeneous solution, the particular solution can be found with substitution into original recurrence relation the ansatz $K^{SE}_{0011100}(\nu) = r(\nu) J_h(\nu)$ and obtaining the recurrence relation for $r(\nu)$
\begin{equation}
r (\nu +1) = r(\nu) + J_h^{-1}(\nu)\Gamma (1-\nu)^2\, .
\end{equation}
This first order recurrence relation is easy to solve and we have
\begin{equation}
r (\nu) = -\sum_{i=0}^{\infty} \frac{3^{2+i-\nu}}{2}\frac{\Gamma (2+i-\nu)^2}{\Gamma (4+2i-2\nu)} = 
-\frac{3^{2-\nu}\Gamma (2-\nu)^2}{2\Gamma (4-2\nu)}~_2F_1\left(
1, 2-\nu ;\frac{5}{2}-\nu ; \frac{3}{4}
\right)\, .
\end{equation}
The general solution is then given by
\begin{equation}
K^{SE}_{0011100}(\nu) = J_h (\nu) r(\nu) + w (\nu) J_h (\nu)\, ,
\end{equation}
where the periodic constant $w (\nu)$ is fixed from the leading term of the expansion at $\nu = 3/2$. Altogether, we have the exact solution,
\begin{multline}
K^{SE}_{0011100}(\nu) = \frac{4\cdot 3^{\nu-3/2}\pi^2\csc (2\pi\nu)}{\Gamma (2\nu -1)} \\
-\frac{3\pi \csc (2\pi\nu)\Gamma^2 (2-\nu)}{\Gamma (4-2\nu)\Gamma (2\nu -1)}~_2F_1\left(
1, 2-\nu; \frac{5}{2}-\nu; \frac{3}{4}
\right)\, .
\end{multline}
Employing the same technique, one can then calculate all other boundary master integrals. The corresponding results can be found in the accompanying {\it Mathematica} notebook.  Expanding the exact results at $d=4-2\ep$ and accounting for obvious $\exp (2\gamma_E\ep)$ factors, we get

\subsubsection*{Heavy-heavy:}
\begin{align}
    a_{1}={}& \frac{1}{\ep ^2}+\frac{2}{\ep }+\left(3+\frac{\pi ^2}{6}\right)+\left(-\frac{2 \zeta_3}{3}+4+\frac{\pi ^2}{3}\right) \ep
   +\left(-\frac{4 \zeta_3}{3}+5+\frac{\pi ^2}{2}+\frac{7 \pi ^4}{360}\right) \ep ^2 \nonumber\\&+\left(-2 \zeta_3-\frac{\pi ^2 \zeta_3}{9}-\frac{2
   \zeta_5}{5}+6+\frac{2 \pi ^2}{3}+\frac{7 \pi ^4}{180}\right) \ep ^3+\left(-\frac{8 \zeta_3}{3}-\frac{2 \pi ^2 \zeta_3}{9} +\frac{2 \zeta_3^2}{9}-\frac{4 \zeta_5}{5}\right. \nonumber\\& \left.+7+\frac{5 \pi ^2}{6}+\frac{7 \pi ^4}{120}+\frac{31 \pi ^6}{15120}\right) \ep ^4+\mathcal{O}(\ep ^5) , \\
    
    a_{2}={}&-\frac{2 \pi }{\ep }+\pi  (4 l_2-6)+\pi  \ep  \left(-14-\frac{2 \pi ^2}{3}-4 l_2^2+12 l_2\right)+\pi  \ep ^2
   \left(\frac{16 \zeta_3}{3} -30-2 \pi ^2 \right. \nonumber\\& \left. +\frac{8 l_2^3}{3}-12 l_2^2+28 l_2+\frac{4}{3} \pi ^2 l_2\right)+\pi  \ep ^3
   \left(16 \zeta_3  -\frac{32}{3} \zeta_3 l_2-62-\frac{14 \pi ^2}{3}-\frac{\pi ^4}{5}-\frac{4 l_2^4}{3}  \right. \nonumber\\& \left. +8 l_2^3-28 l_2^2-\frac{4}{3} \pi ^2 l_2^2 +60 l_2+4 \pi ^2 l_2\right)+\pi  \ep ^4 \left(\frac{112 \zeta_3}{3}+\frac{16 \pi ^2 \zeta_3}{9}+\frac{64 \zeta_5}{5}+\frac{32}{3} \zeta_3 l_2^2 \right.\nonumber\\& \left.  -32 \zeta_3 l_2-126-10 \pi ^2-\frac{3 \pi ^4}{5}+\frac{8 l_2^5}{15}-4
   l_2^4+\frac{56 l_2^3}{3}+\frac{8}{9} \pi ^2 l_2^3  -60 l_2^2-4 \pi ^2 l_2^2+124 l_2 \right.\nonumber\\& \left.+\frac{28}{3} \pi ^2 l_2+\frac{2}{5} \pi ^4 l_2\right)+\mathcal{O}(\ep ^5), \\

    a_{3}={}&\frac{32 \pi ^2 \ep }{3}+\ep ^2 \left(\frac{704 \pi ^2}{9}-\frac{256}{3} \pi ^2 l_2\right)+\ep ^3 \left(\frac{10880 \pi
   ^2}{27}+\frac{112 \pi ^4}{9}+\frac{1024}{3} \pi ^2 l_2^2  -\frac{5632}{9} \pi ^2 l_2\right) \nonumber\\& +\ep ^4 \left(-\frac{1984 \pi ^2 \zeta_3}{9}+\frac{147200 \pi ^2}{81}+\frac{2464 \pi ^4}{27}-\frac{8192}{9} \pi ^2 l_2^3  +\frac{22528}{9} \pi ^2 l_2^2-\frac{87040}{27} \pi ^2
   l_2 \right.\nonumber\\& \left.-\frac{896}{9} \pi ^4 l_2\right)+\mathcal{O}(\ep ^5), \\
   
    a_{4}={}&2\pi ^2+\ep  \left(8 \pi ^2-8 \pi ^2 l_2\right)+\ep ^2 \left(24 \pi ^2+\pi ^4+16 \pi ^2 l_2^2-32 \pi ^2 l_2\right)+\ep ^3
   \left(-\frac{28 \pi ^2 \zeta_3}{3}+64 \pi ^2  \right.\nonumber\\& \left. +4 \pi ^4-\frac{64}{3} \pi ^2 l_2^3+64 \pi ^2 l_2^2-96 \pi ^2 l_2-4 \pi ^4 l_2\right)+\ep ^4 \left(-\frac{112 \pi ^2 \zeta_3}{3}+\frac{112}{3} \pi ^2 \zeta_3 l_2+160 \pi ^2 \right.\nonumber\\& \left.+12 \pi ^4+\frac{5 \pi ^6}{12}+\frac{64}{3} \pi
   ^2 l_2^4-\frac{256}{3} \pi ^2 l_2^3+192 \pi ^2 l_2^2+8 \pi ^4 l_2^2-256 \pi ^2 l_2-16 \pi ^4 l_2\right)+\mathcal{O}(\ep ^5), \\

   a_{5}={}& \frac{3}{2 \ep ^2}+\frac{9}{2 \ep }+\frac{1}{4} \left(8 \sqrt{3} \Im\left(\text{Li}_2\left(\frac{1}{2}-\frac{i
   \sqrt{3}}{2}\right)\right)+42+\pi ^2\right)+ \sum_{n=1}^4a_{1,n}\ep^n +\mathcal{O}(\ep ^5), \\
   
   a_{6}={}& -\frac{2 \pi }{\ep }+\left(-8 \pi -\frac{4 \pi ^2}{\sqrt{3}}+4 \pi  l_2\right)+ \sum_{n=1}^4a_{2,n}\ep^n+\mathcal{O}(\ep ^5) , \\
   
    a_{7}={}&6 \pi ^2+\ep  \left(24 \pi ^2-24 \pi ^2 l_2\right)+\ep ^2 \left(72 \pi ^2+3 \pi ^4+48 \pi
   ^2 l_2^2-96 \pi ^2 l_2\right)+\ep ^3 \left(-28 \pi ^2 \zeta_3+192 \pi
   ^2 \right.\nonumber\\& \left.+12 \pi ^4-64 \pi ^2 l_2^3+192 \pi ^2 l_2^2-288 \pi ^2 l_2-12 \pi ^4 l_2\right)+\ep ^4 \left(-112 \pi ^2 \zeta_3+112 \pi ^2 \zeta_3 l_2+480 \pi ^2 \right.\nonumber\\& \left.+36 \pi ^4+\frac{5 \pi ^6}{4}+64 \pi ^2 l_2^4-256 \pi ^2 l_2^3+576 \pi ^2 l_2^2+24 \pi ^4 l_2^2-768 \pi ^2 l_2-48 \pi ^4 l_2\right)+\mathcal{O}(\ep ^5), \\
   
    a_{8}={}&4 \pi ^2+\ep  \left(16 \pi ^2-16 \pi ^2 l_2\right)+\ep ^2 \left(48 \pi ^2+2 \pi ^4+32 \pi ^2 l_2^2-64 \pi ^2 l_2\right)+\ep ^3
   \left(-\frac{56 \pi ^2 \zeta_3}{3}+128 \pi ^2 \right.\nonumber\\& \left.+8 \pi ^4-\frac{128}{3} \pi ^2 l_2^3+128 \pi ^2 l_2^2-192 \pi ^2 l_2-8 \pi ^4 l_2\right)+\ep ^4 \left(-\frac{224 \pi ^2 \zeta_3}{3}+\frac{224}{3} \pi ^2 \zeta_3 l_2+320 \pi ^2  \right.\nonumber\\& \left. +24 \pi ^4+\frac{5 \pi ^6}{6}+\frac{128}{3} \pi
   ^2 l_2^4-\frac{512}{3} \pi ^2 l_2^3+384 \pi ^2 l_2^2+16 \pi ^4 l_2^2-512 \pi ^2 l_2-32 \pi ^4 l_2\right)+\mathcal{O}(\ep ^5), \\
    
    a_{9}={}& -\frac{4 \pi ^2}{3 \ep }+\left(\frac{16}{3} \pi ^2 l_3-\frac{8 \pi ^2}{3}\right)+ \sum_{n=1}^4a_{3,n}\ep^n+\mathcal{O}(\ep ^5),\\
   
   a_{10}={}& \frac{1}{2} a_9, \\
   
   a_{11}={}& \frac{1}{2 \ep ^2}+\frac{3}{2 \ep }+\left(\frac{7}{2}+\frac{\pi ^2}{4}\right)+\left(-\frac{4 \zeta_3}{3}+\frac{15}{2}+\frac{3 \pi
   ^2}{4}\right) \ep +\left(-4 \zeta_3+\frac{31}{2}+\frac{7 \pi ^2}{4}+\frac{7 \pi ^4}{80}\right) \ep ^2  \nonumber\\& +\left(-\frac{28 \zeta_3}{3}-\frac{2 \pi ^2 \zeta_3}{3}-\frac{16 \zeta_5}{5}+\frac{63}{2}+\frac{15 \pi ^2}{4}+\frac{21 \pi ^4}{80}\right) \ep ^3+\left(-20
   \zeta_3-2 \pi ^2 \zeta_3+\frac{16 \zeta_3^2}{9}  \right.\nonumber\\& \left. -\frac{48 \zeta_5}{5}+\frac{127}{2}+\frac{31 \pi ^2}{4}+\frac{49 \pi ^4}{80}+\frac{869 \pi
   ^6}{30240}\right) \ep ^4+\mathcal{O}(\ep ^5), \\
   
   a_{12}={}&-\frac{2 \pi }{\ep }+ \pi  (4 l_2-8)+\pi 
   \ep  \left(-24-\frac{10 \pi ^2}{3}-4 l_2^2+16 l_2\right)+\pi  \ep ^2 \left(\frac{28 \zeta_3}{3}-64-\frac{40 \pi ^2}{3}+\frac{8 l_2^3}{3}  \right.\nonumber\\& \left. -16 l_2^2+48 l_2+\frac{20}{3} \pi ^2 l_2\right)+\pi  \ep ^3
   \left(\frac{112 \zeta_3}{3}-\frac{56}{3} \zeta_3 l_2-160-40 \pi ^2-\frac{238 \pi ^4}{45}-\frac{4 l_2^4}{3}+\frac{32 l_2^3}{3} \right.\nonumber\\& \left.-48 l_2^2-\frac{20}{3} \pi ^2 l_2^2+128 l_2+\frac{80}{3} \pi ^2 l_2\right)+\pi  \ep ^4 \left(112 \zeta_3+\frac{140 \pi ^2 \zeta_3}{9}+\frac{124 \zeta_5}{5}+\frac{56}{3} \zeta_3 l_2^2 \right.\nonumber\\& \left.-\frac{224}{3} \zeta_3
   l_2   -384-\frac{320 \pi ^2}{3}-\frac{952 \pi ^4}{45}+\frac{8 l_2^5}{15}-\frac{16 l_2^4}{3}+32 l_2^3+\frac{40}{9} \pi ^2 l_2^3-128 l_2^2 \right.\nonumber\\& \left.-\frac{80}{3} \pi ^2 l_2^2+320 l_2  +80 \pi ^2 l_2+\frac{476}{45} \pi ^4 l_2\right)+\mathcal{O}(\ep ^5),
\end{align}
\subsubsection*{Heavy-light:}
\begin{align}
    b_{1}={}& \frac{1}{2 \ep ^2}+\frac{5}{4 \ep }+\left(\frac{11}{8}+\frac{5 \pi ^2}{12}\right)+\left(\frac{11 \zeta_3}{3}-\frac{55}{16}+\frac{25 \pi ^2}{24}\right) \ep +\left(\frac{55 \zeta_3}{6}-\frac{949}{32}+\frac{55 \pi
   ^2}{48}+\frac{101 \pi ^4}{240}\right) \ep ^2 \nonumber\\& +\left(\frac{121 \zeta_3}{12}+\frac{55 \pi ^2 \zeta_3}{18}+\frac{359
   \zeta_5}{5}-\frac{8575}{64}-\frac{275 \pi ^2}{96}+\frac{101 \pi ^4}{96}\right) \ep ^3+\left(-\frac{605 \zeta_3}{24}+\frac{275 \pi ^2 \zeta_3}{36} \right.\nonumber\\& \left. +\frac{121 \zeta_3^2}{9}+\frac{359 \zeta_5}{2}-\frac{64189}{128}-\frac{4745 \pi
   ^2}{192}+\frac{1111 \pi ^4}{960}+\frac{3335 \pi ^6}{6048}\right) \ep ^4+\mathcal{O}(\ep ^5), \\
   
    b_{2}={}&-\frac{1}{\ep ^2}-\frac{4}{3 \ep }+\frac{\pi ^2}{2}+\frac{56}{9}+\left(-\frac{118 \zeta_3}{3}+\frac{1520}{27}+\frac{2 \pi ^2}{3}\right) \ep+\left(-\frac{472 \zeta_3}{9}+\frac{24224}{81}-\frac{28 \pi ^2}{9}\right.\nonumber\\& \left.+\frac{299
   \pi ^4}{120}\right) \ep ^2+\left(\frac{6608 \zeta_3}{27}+\frac{59 \pi ^2 \zeta_3}{3}-\frac{6478 \zeta_5}{5}+\frac{330176}{243}-\frac{760 \pi
   ^2}{27}+\frac{299 \pi ^4}{90}\right) \ep ^3 \nonumber\\& +\left(\frac{179360 \zeta_3}{81}+\frac{236 \pi ^2 \zeta_3}{9}-\frac{6962 \zeta_3^2}{9}-\frac{25912 \zeta_5}{15}+\frac{4213376}{729}-\frac{12112 \pi ^2}{81}-\frac{2093 \pi ^4}{135} \right.\nonumber\\& \left. +\frac{91909 \pi ^6}{15120}\right) \ep
   ^4+\mathcal{O}(\ep ^5), \\
   
    b_{3}={}&\frac{1}{\ep ^2}+\frac{2}{\ep }+\left(3+\frac{\pi ^2}{6}\right)+\left(-\frac{2 \zeta_3}{3}+4+\frac{\pi ^2}{3}\right)
   \ep +\left(-\frac{4 \zeta_3}{3}+5+\frac{\pi ^2}{2}+\frac{7 \pi ^4}{360}\right) \ep ^2+\left(-2 \zeta_3 \right.\nonumber\\& \left. -\frac{\pi
   ^2 \zeta_3}{9}-\frac{2 \zeta_5}{5}+6+\frac{2 \pi ^2}{3}+\frac{7 \pi ^4}{180}\right) \ep ^3+\left(-\frac{8 \zeta_3}{3}-\frac{2 \pi ^2 \zeta_3}{9}+\frac{2 \zeta_3^2}{9}-\frac{4 \zeta_5}{5}+7+\frac{5 \pi ^2}{6}\right.\nonumber\\& \left.+\frac{7 \pi
   ^4}{120}+\frac{31 \pi ^6}{15120}\right) \ep ^4+\mathcal{O}(\ep ^5), \\
   
    b_{4}={}&\frac{32 \pi ^2 \ep }{3}+\ep ^2 \left(\frac{704 \pi ^2}{9}-\frac{256}{3} \pi ^2 l_2\right)+\ep ^3
   \left(\frac{10880 \pi ^2}{27}+\frac{112 \pi ^4}{9}+\frac{1024}{3} \pi ^2 l_2^2-\frac{5632}{9} \pi ^2l_2\right)\nonumber\\&  +\ep ^4 \left(-\frac{1984 \pi ^2 \zeta_3}{9}+\frac{147200 \pi ^2}{81}+\frac{2464 \pi ^4}{27}-\frac{8192}{9}
   \pi ^2 l_2^3+\frac{22528}{9} \pi ^2 l_2^2-\frac{87040}{27} \pi ^2 l_2\right.\nonumber\\& \left.-\frac{896}{9} \pi ^4 l_2\right)+\mathcal{O}(\ep ^5), \\
   
    b_{5}={}&\frac{3}{2 \ep ^2}+\frac{17}{4 \ep }+\left(\frac{59}{8}+\frac{\pi ^2}{4}\right)+\left(-\zeta_3+\frac{65}{16}+\frac{49
   \pi ^2}{24}\right) \ep +\ep ^2 \left(\frac{151 \zeta_3}{6}-\frac{1117}{32}+\frac{475 \pi ^2}{48}+\frac{7 \pi
   ^4}{240}\right.\nonumber\\& \left.-8 \pi ^2 l_2\right)+\ep ^3 \left(192 \text{Li}_4\left(\frac{1}{2}\right)+\frac{2125 \zeta_3}{12}-\frac{\pi ^2 \zeta_3}{6}-\frac{3 \zeta_5}{5}-\frac{13783}{64}+\frac{3745 \pi ^2}{96}-\frac{103 \pi ^4}{96}\right.\nonumber\\& \left.+8 l_2^4+16 \pi ^2 l_2^2-52 \pi ^2 l_2\right)+\ep ^4 \left(1248 \text{Li}_4\left(\frac{1}{2}\right)+1152
   \text{Li}_5\left(\frac{1}{2}\right)+\frac{19255 \zeta_3}{24}-\frac{361 \pi ^2 \zeta_3}{36}\right.\nonumber\\& \left.+\frac{\zeta_3^2}{3}-\frac{9317 \zeta_5}{10}-\frac{114181}{128}+\frac{26563 \pi ^2}{192}-\frac{7073 \pi ^4}{960}+\frac{31 \pi
   ^6}{10080}-\frac{48 l_2^5}{5}+52 l_2^4-32 \pi ^2 l_2^3\right.\nonumber\\& \left.+104 \pi ^2 l_2^2-230 \pi ^2 l_2+\frac{104}{15}
   \pi ^4 l_2\right)+\mathcal{O}(\ep ^5). 
\end{align}
\subsubsection*{Heavy propagator:}
\begin{align}
    c_{2}={}&a_5,\quad c_5=a_6,\quad c_8=\frac{1}{2}a_9 \\
    
    c_{1}={}&\frac{1}{\ep ^2}+\frac{2}{\ep }+\left(3+\frac{\pi ^2}{6}\right)+\left(-\frac{2 \zeta_3}{3}+4+\frac{\pi ^2}{3}\right) \ep
   +\left(-\frac{4 \zeta_3}{3}+5+\frac{\pi ^2}{2}+\frac{7 \pi ^4}{360}\right) \ep ^2+\left(-2 \zeta_3 \right.\nonumber\\&{}\left. -\frac{\pi ^2 \zeta_3}{9}-\frac{2 \zeta_5}{5}+6+\frac{2 \pi ^2}{3}+\frac{7 \pi ^4}{180}\right) \ep ^3+\left(-\frac{8 \zeta_3}{3}-\frac{2 \pi ^2 \zeta_3}{9}+\frac{2 \zeta_3^2}{9}-\frac{4 \zeta_5}{5}+7+\frac{5 \pi ^2}{6}+\frac{7 \pi ^4}{120}\right.\nonumber\\&{}\left.+\frac{31 \pi ^6}{15120}\right) \ep ^4+\mathcal{O}(\ep ^5), \\
   
    c_{3}={}&-\frac{2 \pi }{\ep }+\pi  (4 l_2-6)-\frac{2}{3} \ep  \left(\pi  \left(21+\pi ^2+6 l_2^2-18 l_2\right)\right)+\frac{2}{3} \pi 
   \ep ^2 \left(8 \zeta_3-45-3 \pi ^2+4 l_2^3\right.\nonumber\\&{}\left.-18 l_2^2+42 l_2+2 \pi ^2 l_2\right)-\frac{1}{15} \ep ^3 \left( \pi
   \left(-240 \zeta_3+160 \zeta_3 l_2+930+70 \pi ^2+3 \pi ^4+20 l_2^4\right.\right.\nonumber\\&{}\left.\left.-120 l_2^3+420 l_2^2+20 \pi ^2 l_2^2-900 l_2-60
   \pi ^2 l_2\right)\right)+\frac{1}{45} \pi  \ep ^4 \left(1680 \zeta_3+80 \pi ^2 \zeta_3+576 \zeta_5\right.\nonumber\\&{}\left.+480 \zeta_3 l_2^2-1440
   \zeta_3 l_2-5670-450 \pi ^2-27 \pi ^4+24 l_2^5-180 l_2^4+840 l_2^3+40 \pi ^2 l_2^3\right.\nonumber\\&{}\left.-2700 l_2^2-180 \pi ^2 l_2^2+5580 l_2+420 \pi ^2 l_2+18 \pi ^4 l_2\right)+\mathcal{O}(\ep ^5), \\
   
    c_{4}={}&\frac{32 \pi ^2 \ep }{3}+\ep ^2 \left(\frac{704 \pi ^2}{9}-\frac{256}{3} \pi ^2 l_2\right)+\frac{16}{27} \ep ^3 \left(680 \pi ^2+21
   \pi ^4+576 \pi ^2 l_2^2-1056 \pi ^2 l_2\right)\nonumber\\&{}-\frac{32}{81} \ep ^4 \left(558 \pi ^2 \zeta_3-4600 \pi ^2-231 \pi ^4+2304 \pi ^2 l_2^3-6336 \pi ^2 l_2^2+8160 \pi ^2 l_2\right.\nonumber\\&{}\left.+252 \pi ^4 l_2\right)+\mathcal{O}(\ep ^5), \\
    
    c_6={}&2 \pi ^2+\ep  \left(8 \pi ^2-8 \pi ^2 l_2\right)+\ep ^2 \left(24 \pi ^2+\pi ^4+16 \pi ^2 l_2^2-32 \pi ^2 l_2\right)+\ep ^3
   \left(-\frac{28 \pi ^2 \zeta_3}{3}+64 \pi ^2+4 \pi ^4\right.\nonumber\\&{}\left.-\frac{64}{3} \pi ^2 l_2^3+64 \pi ^2 l_2^2-96 \pi ^2 l_2-4 \pi ^4 l_2\right)+\frac{1}{12} \ep ^4 \left(-448 \pi ^2 \zeta_3+448 \pi ^2 \zeta_3 l_2+1920 \pi ^2+144 \pi ^4\right.\nonumber\\&{}\left.+5 \pi ^6+256 \pi ^2 l_2^4-1024 \pi ^2 l_2^3+2304 \pi ^2 l_2^2+96 \pi ^4 l_2^2-3072 \pi ^2 l_2-192 \pi ^4 l_2\right)+\mathcal{O}(\ep ^5), \\
   
   c_{7}={}& 4 \pi ^2+\ep  \left(16 \pi ^2-16 \pi ^2 l_2\right)+\ep ^2 \left(48 \pi ^2+2 \pi ^4+32 \pi ^2 l_2^2-64 \pi ^2 l_2\right)+\ep ^3 \left(-\frac{56 \pi ^2 \zeta_3}{3}+128 \pi ^2\right.\nonumber\\&{}\left.+8 \pi ^4-\frac{128}{3} \pi ^2 l_2^3+128 \pi ^2 l_2^2-192 \pi ^2 l_2-8 \pi ^4 l_2\right)+\frac{1}{6} \ep ^4 \left(-448 \pi ^2 \zeta_3+448 \pi ^2 \zeta_3 l_2+1920 \pi ^2\right.\nonumber\\&{}\left.+144 \pi ^4+5 \pi ^6+256
   \pi ^2 l_2^4-1024 \pi ^2 l_2^3+2304 \pi ^2 l_2^2+96 \pi ^4 l_2^2-3072 \pi ^2 l_2-192 \pi ^4 l_2\right)+\mathcal{O}(\ep ^5), \\
   
    c_{9}={}& \frac{1}{2 \ep ^2}+\frac{3}{2 \ep }+\frac{1}{4} \left(14+\pi
   ^2\right)+\frac{1}{12} \left(-16 \zeta_3+90+9 \pi ^2\right) \ep+\left(-4 \zeta_3+\frac{31}{2}+\frac{7 \pi ^2}{4}+\frac{7 \pi ^4}{80}\right)
   \ep ^2\nonumber\\&{}+\left(\pi ^2 \left(\frac{15}{4}-\frac{2 \zeta_3}{3}\right)-\frac{28 \zeta_3}{3}-\frac{16
   \zeta_5}{5}+\frac{63}{2}+\frac{21 \pi ^4}{80}\right) \ep ^3+\left(\pi ^2 \left(\frac{31}{4}-2 \zeta_3\right)-20 \zeta_3\right.\nonumber\\&{}\left.+\frac{16 \zeta_3^2}{9}-\frac{48 \zeta_5}{5}+\frac{127}{2}+\frac{49 \pi
   ^4}{80}+\frac{869 \pi ^6}{30240}\right) \ep ^4+\mathcal{O}(\ep ^5), \\
   
    c_{10}={}&-\frac{2 \pi }{\ep }+4 \pi  (l_2-2)+\pi  \ep  \left(-24-\frac{10 \pi ^2}{3}-4 l_2^2+16 l_2\right)+\frac{4}{3} \pi  \ep ^2
   \left(7 \zeta_3-48-10 \pi ^2+2 l_2^3\right.\nonumber\\&{}\left.-12 l_2^2+36 l_2+5 \pi ^2 l_2\right)-\frac{2}{45} \ep ^3 \left(\pi  \left(-840 \zeta_3+420 \zeta_3 l_2+3600+900 \pi ^2+119 \pi ^4+30 l_2^4\right.\right.\nonumber\\&{}\left.\left.-240 l_2^3+1080 l_2^2+150 \pi ^2 l_2^2-2880 l_2-600 \pi ^2
   l_2\right)\right)+\frac{4}{45} \pi  \ep ^4 \left(1260 \zeta_3+175 \pi ^2 \zeta_3+279 \zeta_5\right.\nonumber\\&{}\left.+210 \zeta_3 l_2^2-840 \zeta_3
   l_2-4320-1200 \pi ^2-238 \pi ^4+6 l_2^5-60 l_2^4+360 l_2^3+50 \pi ^2 l_2^3-1440 l_2^2\right.\nonumber\\&{}\left.-300 \pi ^2 l_2^2+3600 l_2+900 \pi ^2 l_2+119 \pi ^4 l_2\right)+\mathcal{O}(\ep ^5). 
\end{align}
where $\zeta_n=\sum_{n=1}^{\infty}\frac{1}{n^s}$ is the Riemann zeta function, $l_k=\ln{k}$. Here $a_i$ constants correspond to $\# i$ boundary masters in Fig.~\ref{fig:mastersHH-boundary}, $b_i$ constants  to $\# i$ masters in Fig.~\ref{fig:mastersHL-boundary} and $c_i$ constants to  $\# i$ masters in Fig.~\ref{fig:mastersSE-boundary}. The constants $a_{i,j}$ are given in an accompanying {\it Mathematica} notebook. Note, that some of their analytical expressions contain derivatives of hypergeometric functions. The latter can be considered as new elliptic constants\footnote{The differential equation system for heavy-heavy vertex was reduced to $\ep$-form before. Still elliptics is present in its boundary conditions.}, which can be evaluated with very high precision using their triangle sum representation and \texttt{SummerTime} package \cite{SummerTime}, see accompanying  {\it Mathematica} notebook for details.

\subsection{Results}
\label{scn:appDE}
The complete set of results for the canonical MIs up to $\mathcal{O}(\ep^3)$ for the heavy-heavy vertex and $\mathcal{O}(\ep^2)$ for the heavy-light vertex are given in an accompanying {\it Mathematica} notebook. The same notebook also contains results for IBP MIs up to $\mathcal{O}(\ep^2)$ in the case of heavy-heavy and heavy-light vertexes and up to $\mathcal{O}(\ep^2,w)$ in the case of the heavy propagator. Moreover, we provide exact results for the boundary integrals and associated asymptotic coefficient vectors, $\boldsymbol{c}$, as well as the set of matrices, $\lbrace \mathbb{M},\mathbb{T},\mathbb{L}\rbrace$ for all cases. The interested reader can employ these to reproduce our results. Due to the size of the expressions for the canonical masters, we only present terms up to an appropriate order. Also, in the case of the heavy propagator, we only present results for IBP masters.  We omit overall pre-factors of mass scale as these can be determined by inspection. 

\subsubsection*{Heavy-heavy: }
\begin{align}
    \tilde{A}_1={}&\frac{1}{\ep ^2}+\frac{2}{\ep }+\left(3+\frac{\pi ^2}{6}\right)+\frac{1}{3} \left(-2 \zeta_3+12+\pi ^2\right)
   \ep +\frac{1}{360} \left(-480 \zeta_3+1800+180 \pi ^2\nonumber\right.\\&{}\left.+7 \pi ^4\right) \ep ^2+\left(-\frac{1}{9} \pi ^2
   (\zeta_3-6)-2 \zeta_3-\frac{2 \zeta_5}{5}+6+\frac{7 \pi ^4}{180}\right) \ep ^3+\mathcal{O}\left(\ep ^4\right), \\
   
    \tilde{A}_2={}&\frac{35 \pi }{3}+\frac{35}{18} \pi  \ep  (44-12 l_2)+\frac{35}{54} \pi  \ep ^2 \left(680+6 \pi ^2+36 l_2^2-264 l_2\right)\nonumber\\&{}-\frac{35}{81} \ep ^3 \left(4 \pi  \left(18 \zeta_3-1150+9 l_2^3-99 l_2^2+510 l_2\right)+\pi ^3 (18 l_2-66)\right)+\mathcal{O}\left(\ep ^4\right), \\
   
    \tilde{A}_3={}&\frac{32 \pi ^2 \ep }{3}+\frac{64}{9} \pi ^2 \ep ^2 (11-12 l_2)+\frac{16}{27} \ep ^3 \left(21 \pi
   ^4+8 \pi ^2 \left(85+72 l_2^2-132 l_2\right)\right)+\mathcal{O}\left(\ep ^4\right), \\
    
    \tilde{A}_4={}&-\frac{2 \pi }{\ep }+4 \pi  (l_2-2)-\frac{2}{3} \ep  \left(\pi  \left(36+5 \pi ^2+6 l_2^2-24 l_2)\right)\right)+\frac{4}{3} \pi  \ep ^2 \left(7 \zeta_3-48+2 l_2^3\nonumber\right.\\&{}\left.-12 l_2^2+5 \pi ^2 (l_2-2)+36 l_2\right)-\frac{2}{45} \ep ^3 \left(\pi  \left(30 \left(-28 \zeta_3+2 (7 \zeta_3-48) l_2+120\nonumber\right.\right.\right.\\&{}\left.\left.\left.+l_2^4-8 l_2^3+36 l_2^2\right)+119 \pi ^4+150 \pi ^2 \left(6+l_2^2-4 l_2\right)\right)\right)+\mathcal{O}\left(\ep ^4\right), \\
    
    \tilde{A}_5={}&-\frac{i (G(-i;\beta )-G(i;\beta ))}{18 \ep }-\frac{1}{9} i (G(-i;\beta )-G(i;\beta
   )+G(0,-i;\beta ) \nonumber\\&{}-G(0,i;\beta )-G(-i,-i;\beta )+G(-i,i;\beta )-G(i,-i;\beta )\nonumber\\&{}+G(i,i;\beta
   ))+\mathcal{O}\left(\ep ^1\right), \\
    
    \tilde{A}_6={}&-20 i \ep  (\pi  G(-i;\beta )-\pi  G(i;\beta ))+\ep ^2 \left(40 \pi ^2 G(-1;\beta )-40 \pi ^2
   G(1;\beta )\nonumber\right.\\&{}\left.+80 i \pi  G(-1,-i;\beta )-80 i \pi  G(-1,i;\beta )-40 i \pi  G(0,-i;\beta )+40 i \pi 
   G(0,i;\beta )\nonumber\right.\\&{}\left.-40 i \pi  G(-i,-i;\beta )+40 i \pi  G(-i,i;\beta )-40 i \pi  G(i,-i;\beta )+40 i \pi 
   G(i,i;\beta )\nonumber\right.\\&{}\left.+80 i \pi  G(1,-i;\beta )-80 i \pi  G(1,i;\beta )+\frac{10}{3} i \pi  (12 l_2-44)
   G(-i;\beta )\nonumber\right.\\&{}\left.-\frac{10}{3} i \pi  (12 l_2-44) G(i;\beta )\right)+\mathcal{O}\left(\ep ^3\right), \\
    
    \tilde{A}_7={}&-\frac{1}{9 \ep ^2}-\frac{2}{9 \ep }+\frac{1}{54} \left(-48 G(-i,-i;\beta )+48 G(-i,i;\beta )+48
   G(i,-i;\beta )\nonumber\right.\\&{}\left.-48 G(i,i;\beta )-7 \pi ^2-18\right)+\mathcal{O}\left(\ep ^1\right), \\
    
    \tilde{A}_8={}&\frac{i (G(-i;\beta )-G(i;\beta ))}{3 \ep }+\frac{2}{3} i (G(-i;\beta )-G(i;\beta ))+\ep 
   \frac{1}{18} i \left(18+\pi ^2\right)\left( G(-i;\beta )\nonumber\right.\\&{}\left.- G(i;\beta
   )\right)+\mathcal{O}\left(\ep ^2\right), \\
    
    \tilde{A}_9={}&\frac{i (G(-i;\beta )-G(i;\beta ))}{\ep }+2 i (G(-i;\beta )-G(i;\beta ))+\ep  \frac{1}{6}
   i \left(18+\pi ^2\right)\left( G(-i;\beta )\nonumber\right.\\&{}\left.-G(i;\beta )\right)+\mathcal{O}\left(\ep
   ^2\right), \\
    
    \tilde{A}_{10}={}&\ep  \left(5 i \pi  G(-i;\beta )-5 i \pi  G(i;\beta )+\frac{5 \pi }{2}\right)+\ep ^2 \left(-10 i \pi 
   G(0,-i;\beta )+10 i \pi  G(0,i;\beta )\nonumber\right.\\&{}\left.+10 i \pi  G(-i,-i;\beta )-10 i \pi  G(-i,i;\beta )+10 i \pi 
   G(i,-i;\beta )-10 i \pi  G(i,i;\beta )\nonumber\right.\\&{}\left.-\frac{5}{6} i \pi  (12 l_2-44) G(-i;\beta )+\frac{5}{6} i \pi 
   (12 l_2-44) G(i;\beta )+\pi  \left(\frac{95}{6}-5 l_2\right)\right)+\mathcal{O}\left(\ep ^3\right), \\
    
    \tilde{A}_{11}={}&\frac{35 \pi  \ep }{6}+\frac{35}{18} \ep ^2 (12 i \pi  G(0,-i;\beta )-12 i \pi  G(0,i;\beta )+25 \pi -6
   \pi  l_2)\nonumber\\&{}+\frac{35}{54} \ep ^3 \left(-72 \pi ^2 G(0,-1;\beta )+264 i \pi  G(0,-i;\beta )-264 i \pi 
   G(0,i;\beta )\nonumber\right.\\&{}\left.+72 \pi ^2 G(0,1;\beta )-144 i \pi  G(0,-1,-i;\beta )+144 i \pi  G(0,-1,i;\beta )\nonumber\right.\\&{}\left.+72 i
   \pi  G(0,0,-i;\beta )-72 i \pi  G(0,0,i;\beta )+72 i \pi  G(0,-i,-i;\beta )\nonumber\right.\\&{}\left.-72 i \pi  G(0,-i,i;\beta
   )+72 i \pi  G(0,i,-i;\beta )-72 i \pi  G(0,i,i;\beta )\nonumber\right.\\&{}\left.-144 i \pi  G(0,1,-i;\beta )+144 i \pi 
   G(0,1,i;\beta )-72 i \pi  l_2 G(0,-i;\beta )\nonumber\right.\\&{}\left.+72 i \pi  l_2 G(0,i;\beta )+3 \pi ^3+415 \pi +18
   \pi  l_2^2-150 \pi  l_2\right)+\mathcal{O}\left(\ep ^4\right), \\
    
    \tilde{A}_{12}={}&(G(-i,-i;\beta )-G(-i,i;\beta )-G(i,-i;\beta )+G(i,i;\beta ))+2 \ep  (G(-i,-i;\beta
   )\nonumber\\&{}-G(-i,i;\beta )-G(i,-i;\beta )+G(i,i;\beta )-G(-i,0,-i;\beta )+G(-i,0,i;\beta
   )\nonumber\\&{}+G(-i,-i,-i;\beta )-G(-i,-i,i;\beta )+G(-i,i,-i;\beta )-G(-i,i,i;\beta )\nonumber\\&{}+G(i,0,-i;\beta
   )-G(i,0,i;\beta )-G(i,-i,-i;\beta )+G(i,-i,i;\beta )\nonumber\\&{}-G(i,i,-i;\beta )+G(i,i,i;\beta
   ))+\mathcal{O}\left(\ep ^2\right), \\
    
    \tilde{A}_{13}={}&\frac{1}{2 \ep ^2}+\frac{1}{\ep }+\frac{1}{12} \left(8 \sqrt{3} \Im\left(\text{Li}_2\left(\frac{1}{2}-\frac{i
   \sqrt{3}}{2}\right)\right)+18+\pi ^2\right)+\frac{1}{12} \ep  \left(4 a_{1,1}\nonumber\right.\\&{}\left.-8 \sqrt{3}
   \Im\left(\text{Li}_2\left(\frac{1}{2}-\frac{i \sqrt{3}}{2}\right)\right)-\pi ^2-66\right)+\mathcal{O}\left(\ep ^2\right), \\
    
    \tilde{A}_{14}={}&\frac{35 \pi }{9}+\frac{10}{27} \pi  \ep  \left(77+\left(7 \sqrt{3}-12\right) \pi -21 l_2\right)+\frac{5}{162}
   \ep ^2 \left(-63 a_{2,1}+8 \pi ^2 \left(-195+35 \sqrt{3}\nonumber\right.\right.\\&{}\left.\left.+144 l_2\right)-56 \pi  (15 l_2-58)\right)+\mathcal{O}\left(\ep ^3\right), \\
    
    \tilde{A}_{15}={}&\frac{4 \pi ^2}{9}-\frac{8}{9} \ep  \left(\pi ^2 (2 l_2-1)\right)+\frac{2}{9} \pi ^2 \ep ^2 \left(6+\pi
   ^2+16 l_2^2-16 l_2\right)-\frac{4}{27} \ep ^3 \left(2 \pi ^2 \left(7 \zeta_3-6+16 l_2^3\nonumber\right.\right.\\&{}\left.\left.-24 l_2^2+18 l_2\right)+\pi ^4 (6 l_2-3)\right)+\mathcal{O}\left(\ep ^4\right), \\
    
    \tilde{A}_{16}={}&-\frac{2 \pi ^2}{3 \ep }+\frac{4}{3} \pi ^2 (2 l_3-1)+\frac{1}{2}\ep  a_{3,1} + \frac{1}{2}\ep ^2
   a_{3,2} + \frac{1}{2}\ep ^3 a_{3,3} +\mathcal{O}\left(\ep ^4\right), \\
    
    \tilde{A}_{17}={}&-\frac{1}{6} i \ep  \left(\pi ^2 (G(-i;\beta )-G(i;\beta ))+6 (G(-i,-i,-i;\beta )-G(-i,-i,i;\beta )\nonumber\right.\\&{}\left.-G(-i,i,-i;\beta
   )+G(-i,i,i;\beta )-G(i,-i,-i;\beta )+G(i,-i,i;\beta )+G(i,i,-i;\beta )\nonumber\right.\\&{}\left.-G(i,i,i;\beta ))\right)+\mathcal{O}(\ep^2), \\
   
    \tilde{A}_{18}={}&-\frac{35}{3} \pi  \ep ^2 \left(2 \pi  G\left(-\sqrt{3};\beta \right)-2 \pi  G\left(\sqrt{3};\beta \right)-3 i
   \left(G(-i,-i;\beta )-G(-i,i;\beta )\nonumber\right.\right.\\&{}\left.\left.+G(i,-i;\beta ) -G(i,i;\beta )-G\left(-\sqrt{3},-i;\beta
   \right)+G\left(-\sqrt{3},i;\beta \right)-G\left(\sqrt{3},-i;\beta \right)\nonumber\right.\right.\\&{}\left.\left.+G\left(\sqrt{3},i;\beta
   \right)\right)\right)+\mathcal{O}(\ep^3), \\
   
    \tilde{A}_{19}={}&\frac{\ep}{3 \sqrt{3}}  \left(\pi ^2 \left(-G\left(-\frac{1}{\sqrt{3}};\beta \right)+G\left(\frac{1}{\sqrt{3}};\beta \right)+4G\left(-\sqrt{3};\beta \right)-4 G\left(\sqrt{3};\beta \right)\right) \nonumber\right.\\&{}\left.+6 \left(-3G\left(-\frac{1}{\sqrt{3}},-i,-i;\beta \right)+3 \left(G\left(-\frac{1}{\sqrt{3}},-i,i;\beta
   \right)+G\left(-\frac{1}{\sqrt{3}},i,-i;\beta \right) \nonumber\right.\right.\right.\\&{}\left.\left.\left.-G\left(-\frac{1}{\sqrt{3}},i,i;\beta
   \right)+G\left(\frac{1}{\sqrt{3}},-i,-i;\beta \right)-G\left(\frac{1}{\sqrt{3}},-i,i;\beta
   \right)-G\left(\frac{1}{\sqrt{3}},i,-i;\beta \right)\nonumber\right.\right.\right.\\&{}\left.\left.\left.+G\left(\frac{1}{\sqrt{3}},i,i;\beta
   \right)+G\left(-\sqrt{3},-i,-i;\beta \right)-G\left(-\sqrt{3},-i,i;\beta
   \right)-G\left(-\sqrt{3},i,-i;\beta \right)\nonumber\right.\right.\right.\\&{}\left.\left.\left.+G\left(-\sqrt{3},i,i;\beta
   \right)-G\left(\sqrt{3},-i,-i;\beta \right)+G\left(\sqrt{3},-i,i;\beta
   \right)+G\left(\sqrt{3},i,-i;\beta \right)\nonumber\right.\right.\right.\\&{}\left.\left.\left.-G\left(\sqrt{3},i,i;\beta \right)\right)+2 i
   \Im\left(\text{Li}_2\left(\frac{1}{2}-\frac{i \sqrt{3}}{2}\right)\right) (G(-i;\beta )-G(i;\beta ))\right)\right)+\mathcal{O}(\ep^2), \\

    \tilde{A}_{20}={}&-6 G(-i,-i;\beta )+6 G(-i,i;\beta )+6 G(i,-i;\beta )-6 G(i,i;\beta )-\frac{17 \pi ^2}{12}+\mathcal{O}(\ep), \\
    
    \tilde{A}_{21}={}&\frac{i (G(-i;\beta )-G(i;\beta ))}{6 \ep }+\frac{1}{3} i (G(-i;\beta )-G(i;\beta ))\nonumber\\&{}+\frac{1}{36} i
   \ep  \left(24 \sqrt{3} \Im\left(\text{Li}_2\left(\frac{1}{2}-\frac{i \sqrt{3}}{2}\right)\right)+18+\pi ^2\right)
   (G(-i;\beta )-G(i;\beta ))\nonumber\\&{}-\frac{1}{36} i \ep ^2 (G(-i;\beta )-G(i;\beta )) \left(-12 a_{1,1}+24
   \sqrt{3} \Im\left(\text{Li}_2\left(\frac{1}{2}-\frac{i \sqrt{3}}{2}\right)\right)-8 \zeta_3\nonumber\right.\\&{}\left.+7 \pi
   ^2+246\right)+\mathcal{O}\left(\ep ^3\right), \\
    
    \tilde{A}_{22}={}&\frac{35}{3} i \ep  (\pi  G(-i;\beta )-\pi  G(i;\beta ))+\ep ^2 \left(\frac{35}{18} i \pi  (12 l_2-44) G(i;\beta )\nonumber\right.\\&{}\left.-\frac{35}{18} i \pi  (12 l_2-44) G(-i;\beta )\right)+\ep ^3 \left(\frac{35}{54}
   i \pi  \left(680+6 \pi ^2+36 l_2^2-264 l_2\right) G(-i;\beta )\nonumber\right.\\&{}\left.-\frac{35}{54} i \pi  \left(680+6 \pi
   ^2+36 l_2^2-264 l_2\right) G(i;\beta )\right)+\mathcal{O}\left(\ep ^4\right), \\
    
    \tilde{A}_{23}={}&-\frac{35}{6} \pi  \ep ^3 \left(2 \pi  G\left(0,-\sqrt{3};\beta \right)-2 \pi  G\left(0,\sqrt{3};\beta \right)-3 i
   \left(G(0,-i,-i;\beta )-G(0,-i,i;\beta )\nonumber\right.\right.\\&{}\left.\left.+G(0,i,-i;\beta )-G(0,i,i;\beta )-G\left(0,-\sqrt{3},-i;\beta
   \right)+G\left(0,-\sqrt{3},i;\beta \right)\nonumber\right.\right.\\&{}\left.\left.-G\left(0,\sqrt{3},-i;\beta \right)+G\left(0,\sqrt{3},i;\beta
   \right)\right)\right)+\mathcal{O}(\ep^4), \\
   
    \tilde{A}_{24}={}&\frac{1}{2 \ep ^2}+\frac{3}{2 \ep }+\frac{1}{4} \left(14+\pi ^2\right)+\frac{1}{12} \left(-16 \zeta_3+90+9
   \pi ^2\right) \ep +\frac{1}{80} \left(-320 \zeta_3+1240+140 \pi ^2\right.\nonumber\\&{}\left.+7 \pi ^4\right) \ep ^2+\frac{1}{240}
   \left(-2240 \zeta_3-160 \pi ^2 \zeta_3-768 \zeta_5+7560+900 \pi ^2+63 \pi ^4\right) \ep ^3+\mathcal{O}\left(\ep
   ^4\right), \\
    
    \tilde{A}_{25}={}&2 \pi ^2-8 \ep  \left(\pi ^2 (l_2-1)\right)+\ep ^2 \left(\pi ^4+8 \pi ^2 \left(3+2 l_2^2-4l_2\right)\right)-\frac{4}{3} \ep ^3 \left(\pi ^2 \left(7 \zeta_3-48+16 l_2^3\nonumber\right.\right.\\&{}\left.\left.-48 l_2^2+72 l_2+\pi ^2 (3 l_2-3)\right)\right)+\mathcal{O}\left(\ep ^4\right), \\
    
    \tilde{A}_{26}={}&\frac{i (G(-i;\beta )-G(i;\beta ))}{2 \ep }+\frac{3}{2} i (G(-i;\beta )-G(i;\beta ))+\ep \frac{1}{4} i \left(14+\pi ^2\right) 
   \left(G(-i;\beta )\nonumber\right.\\&{}\left.- G(i;\beta
   )\right)+\ep ^2\frac{1}{12} i \left(-16 \zeta_3+90+9 \pi ^2\right) \left( G(-i;\beta )- G(i;\beta )\right)+\mathcal{O}\left(\ep ^3\right).
\end{align}
\subsubsection*{Heavy-light: }
\begin{align}
        \tilde{B}_1={}&\frac{1}{\ep ^2}+\frac{2}{\ep }+\left(3+\frac{\pi ^2}{6}\right)+\frac{1}{3} \left(-2 \zeta_3+12+\pi ^2\right)
   \ep +\frac{1}{360} \left(-480 \zeta_3+1800\nonumber\right.\\&{}\left.+180 \pi ^2+7 \pi ^4\right) \ep ^2+\mathcal{O}\left(\ep ^3\right), \\
   
    \tilde{B}_2={}&\frac{7}{60 \ep ^2}+\frac{7}{30 \ep }+\left(\frac{7}{20}+\frac{7 \pi ^2}{72}\right)+\frac{7}{180} \left(22
   \zeta_3+12+5 \pi ^2\right) \ep \nonumber\\&{} +\frac{7 \left(1760 \zeta_3+600+300 \pi ^2+101 \pi ^4\right) \ep
   ^2}{7200}+\mathcal{O}\left(\ep ^3\right), \\
    
    \tilde{B}_3={}&\frac{2}{105 \ep ^2}+\frac{4}{105 \ep }+\frac{1}{105} \left(-16 G(-i,-i;\beta )+16 G(-i,i;\beta )+16
   G(i,-i;\beta )\nonumber\right.\\&{}\left.-16 G(i,i;\beta )-\pi ^2+6\right)-\frac{2}{315} \ep  \left(48 G(-i,-i;\beta )-48
   G(-i,i;\beta )\nonumber\right.\\&{}\left.-48 G(i,-i;\beta )+48 G(i,i;\beta )-144 G(-i,0,-i;\beta )+144 G(-i,0,i;\beta )\nonumber\right.\\&{}\left.+144
   G(-i,-i,-i;\beta )-144 G(-i,-i,i;\beta )+144 G(-i,i,-i;\beta )\nonumber\right.\\&{}\left.-144 G(-i,i,i;\beta )+144
   G(i,0,-i;\beta )-144 G(i,0,i;\beta )-144 G(i,-i,-i;\beta )\nonumber\right.\\&{}\left.+144 G(i,-i,i;\beta )-144 G(i,i,-i;\beta
   )+144 G(i,i,i;\beta )-118 \zeta_3\nonumber\right.\\&{}\left.+3 \pi ^2-12\right)+\mathcal{O}\left(\ep ^2\right), \\
    
    \tilde{B}_4={}&-\frac{8 i (G(-i;\beta )-G(i;\beta ))}{105 \ep }-\frac{16}{105} i (G(-i;\beta )-G(i;\beta )-3
   G(0,-i;\beta )\nonumber\\&{}+3 G(0,i;\beta )+3 G(-i,-i;\beta )-3 G(-i,i;\beta )+3 G(i,-i;\beta )\nonumber\\&{}-3
   G(i,i;\beta ))+\mathcal{O}\left(\ep ^1\right), \\
    
    \tilde{B}_5={}&-\frac{64 \pi ^2}{105}+\frac{128}{105} \pi ^2 \ep  (4 l_2-1)-\frac{32}{315} \ep ^2 \left(\pi ^2 \left(18+7
   \pi ^2+192 l_2^2-96 l_2\right)\right)+\mathcal{O}\left(\ep ^3\right), \\
    
    \tilde{B}_6={}&\frac{4 i (G(-i;\beta )-G(i;\beta ))}{3 \ep }+\frac{8}{3} i (G(-i;\beta )-G(i;\beta
   )-G(0,-i;\beta )+G(0,i;\beta )\nonumber\\&{}+G(-i,-i;\beta )-G(-i,i;\beta )+G(i,-i;\beta )-G(i,i;\beta
   ))+\mathcal{O}\left(\ep ^1\right), \\
    
    \tilde{B}_7={}&\frac{14}{45} i \ep  \left(\pi ^2 (G(-i;\beta )-G(i;\beta ))+6 (G(-i,-i,-i;\beta )-G(-i,-i,i;\beta )\nonumber\right.\\&{}\left.-G(-i,i,-i;\beta
   ).+G(-i,i,i;\beta )-G(i,-i,-i;\beta )+G(i,-i,i;\beta )+G(i,i,-i;\beta )\nonumber\right.\\&{}\left.-G(i,i,i;\beta ))\right)+\mathcal{O}(\ep^2), \\
   
    \tilde{B}_8={}&-\frac{4 i (G(-i;\beta )-G(i;\beta ))}{15 \ep }-\frac{8}{15} i (G(-i;\beta )-G(i;\beta )-2
   G(0,-i;\beta )\nonumber\\&{}+2 G(0,i;\beta )+2 G(-i,-i;\beta )-2 G(-i,i;\beta )+2 G(i,-i;\beta )\nonumber\\&{}-2
   G(i,i;\beta ))+\mathcal{O}\left(\ep ^1\right), \\
    
    \tilde{B}_9={}&\frac{1}{189} i \ep  \left(\pi ^2 (G(-i;\beta )-G(i;\beta ))+6 (G(-i,-i,-i;\beta )-G(-i,-i,i;\beta )\nonumber\right.\\&{}\left.-G(-i,i,-i;\beta
   )+G(-i,i,i;\beta )-G(i,-i,-i;\beta )+G(i,-i,i;\beta )+G(i,i,-i;\beta )\nonumber\right.\\&{}\left.-G(i,i,i;\beta ))\right)+\mathcal{O}(\ep^0), \\
   
    \tilde{B}_{10}={}&\frac{1}{441} \left(2 G(-i,-i;\beta )-2 G(-i,i;\beta )-2 G(i,-i;\beta )+2 G(i,i;\beta )+\pi
   ^2\right)\nonumber\\&{}+\frac{2}{441} \ep  \left(\pi ^2 G(-i;\beta )+\pi ^2 G(i;\beta )+2 G(-i,-i;\beta )-2
   G(-i,i;\beta )\nonumber\right.\\&{}\left.-2 G(i,-i;\beta )+2 G(i,i;\beta )-4 G(0,-i,-i;\beta )+4 G(0,-i,i;\beta )\nonumber\right.\\&{}\left.+4
   G(0,i,-i;\beta )-4 G(0,i,i;\beta )-6 G(-i,0,-i;\beta )+6 G(-i,0,i;\beta )\nonumber\right.\\&{}\left.+10 G(-i,-i,-i;\beta )-10
   G(-i,-i,i;\beta )+2 G(-i,i,-i;\beta )-2 G(-i,i,i;\beta )\nonumber\right.\\&{}\left.+6 G(i,0,-i;\beta )-6 G(i,0,i;\beta )-2
   G(i,-i,-i;\beta )+2 G(i,-i,i;\beta )\nonumber\right.\\&{}\left.-10 G(i,i,-i;\beta )+10 G(i,i,i;\beta )+\pi ^2-4 \pi ^2 l_2\right)+\mathcal{O}\left(\ep ^2\right), \\
    
    \tilde{B}_{11}={}&-\frac{8}{9} \left(2 G(-i,-i;\beta )-2 G(-i,i;\beta )-2 G(i,-i;\beta )+2 G(i,i;\beta )+\pi
   ^2\right)\nonumber\\&{}-\frac{16}{9} \ep  \left(\pi ^2 G(-i;\beta )+\pi ^2 G(i;\beta )+2 G(-i,-i;\beta )-2
   G(-i,i;\beta )\nonumber\right.\\&{}\left.-2 G(i,-i;\beta )+2 G(i,i;\beta )-4 G(0,-i,-i;\beta )+4 G(0,-i,i;\beta )\nonumber\right.\\&{}\left.+4
   G(0,i,-i;\beta )-4 G(0,i,i;\beta )-2 G(-i,0,-i;\beta )+2 G(-i,0,i;\beta )\nonumber\right.\\&{}\left.+6 G(-i,-i,-i;\beta )-6
   G(-i,-i,i;\beta )-2 G(-i,i,-i;\beta )+2 G(-i,i,i;\beta )\nonumber\right.\\&{}\left.+2 G(i,0,-i;\beta )-2 G(i,0,i;\beta )+2
   G(i,-i,-i;\beta )-2 G(i,-i,i;\beta )\nonumber\right.\\&{}\left.-6 G(i,i,-i;\beta )+6 G(i,i,i;\beta )+\pi ^2-4 \pi ^2l_2\right)+\mathcal{O}\left(\ep ^2\right), \\
    
    \tilde{B}_{12}={}&-\frac{2}{11} (G(-i,-i;\beta )-G(-i,i;\beta )-G(i,-i;\beta )+G(i,i;\beta ))\nonumber\\&{}-\frac{4}{11} \ep 
   (G(-i,-i;\beta )-G(-i,i;\beta )-G(i,-i;\beta )+G(i,i;\beta )\nonumber\\&{}-2 G(0,-i,-i;\beta )+2
   G(0,-i,i;\beta )+2 G(0,i,-i;\beta )-2 G(0,i,i;\beta )\nonumber\\&{}-3 G(-i,0,-i;\beta )+3 G(-i,0,i;\beta )+5
   G(-i,-i,-i;\beta )-5 G(-i,-i,i;\beta )\nonumber\\&{}+G(-i,i,-i;\beta )-G(-i,i,i;\beta )+3 G(i,0,-i;\beta )-3
   G(i,0,i;\beta )\nonumber\\&{}-G(i,-i,-i;\beta )+G(i,-i,i;\beta )-5 G(i,i,-i;\beta )+5 G(i,i,i;\beta
   ))+\mathcal{O}\left(\ep ^2\right), \\
    
    \tilde{B}_{13}={}&i \ep ^2 (2 G(-i,0,-i,-i;\beta )-2 (G(-i,0,-i,i;\beta )+G(-i,0,i,-i;\beta )-G(-i,0,i,i;\beta )\nonumber\\&{}+G(-i,-i,0,-i;\beta
   )-G(-i,-i,0,i;\beta )-2 G(-i,-i,i,-i;\beta )+2 G(-i,-i,i,i;\beta )\nonumber\\&{}-G(-i,i,0,-i;\beta )+G(-i,i,0,i;\beta )+2
   G(-i,i,-i,-i;\beta )-2 G(-i,i,-i,i;\beta )\nonumber\\&{}+G(i,0,-i,-i;\beta )-G(i,0,-i,i;\beta )-G(i,0,i,-i;\beta )+G(i,0,i,i;\beta
   )\nonumber\\&{}-G(i,-i,0,-i;\beta )+G(i,-i,0,i;\beta )+2 G(i,-i,i,-i;\beta )-2 G(i,-i,i,i;\beta )\nonumber\\&{}+G(i,i,0,-i;\beta )-G(i,i,0,i;\beta )-2
   G(i,i,-i,-i;\beta )+2 G(i,i,-i,i;\beta ))\nonumber\\&{}+3 \zeta_3 (G(i;\beta )-G(-i;\beta )))+\mathcal{O}(\ep^3), \\
   
    \tilde{B}_{14}={}&\frac{1}{4} (G(-i,-i;\beta )-G(-i,i;\beta )-G(i,-i;\beta )+G(i,i;\beta ))\nonumber\\&{}+\frac{1}{2} \ep 
   (G(-i,-i;\beta )-G(-i,i;\beta )-G(i,-i;\beta )+G(i,i;\beta )\nonumber\\&{}-2 G(0,-i,-i;\beta )+2
   G(0,-i,i;\beta )+2 G(0,i,-i;\beta )-2 G(0,i,i;\beta )\nonumber\\&{}-G(-i,0,-i;\beta )+G(-i,0,i;\beta )+3
   G(-i,-i,-i;\beta )-3 G(-i,-i,i;\beta )\nonumber\\&{}-G(-i,i,-i;\beta )+G(-i,i,i;\beta )+G(i,0,-i;\beta
   )-G(i,0,i;\beta )\nonumber\\&{}+G(i,-i,-i;\beta )-G(i,-i,i;\beta )-3 G(i,i,-i;\beta )+3 G(i,i,i;\beta
   ))+\mathcal{O}\left(\ep ^2\right), \\
    
    \tilde{B}_{15}={}&\frac{32}{9} (G(-i,-i;\beta )-G(-i,i;\beta )-G(i,-i;\beta )+G(i,i;\beta ))\nonumber\\&{}+\frac{64}{9} \ep 
   (G(-i,-i;\beta )-G(-i,i;\beta )-G(i,-i;\beta )+G(i,i;\beta )\nonumber\\&{}-2 G(0,-i,-i;\beta )+2
   G(0,-i,i;\beta )+2 G(0,i,-i;\beta )-2 G(0,i,i;\beta )\nonumber\\&{}-G(-i,0,-i;\beta )+G(-i,0,i;\beta )+3
   G(-i,-i,-i;\beta )-3 G(-i,-i,i;\beta )\nonumber\\&{}-G(-i,i,-i;\beta )+G(-i,i,i;\beta )+G(i,0,-i;\beta
   )-G(i,0,i;\beta )\nonumber\\&{}+G(i,-i,-i;\beta )-G(i,-i,i;\beta )-3 G(i,i,-i;\beta )+3 G(i,i,i;\beta
   ))+\mathcal{O}\left(\ep ^2\right), \\
    
    \tilde{B}_{16}={}&\frac{19}{60 \ep ^2}+\frac{19}{30 \ep }+\frac{1}{360} \left(342-13 \pi ^2\right)+\frac{1}{180} \ep 
   \left(-374 \zeta_3+228+\pi ^2 (96 l_2-13)\right)+\mathcal{O}\left(\ep ^2\right), \\
    
    \tilde{B}_{17}={}&\frac{1}{15} \left(-8 G(-i,-i;\beta )+8 G(-i,i;\beta )+8 G(i,-i;\beta )-8 G(i,i;\beta )-\pi
   ^2\right)\nonumber\\&{}+\frac{1}{15} \ep  \left(4 \pi ^2 G(-1;\beta )-8 \pi ^2 G(-i;\beta )-8 \pi ^2 G(i;\beta )+4
   \pi ^2 G(1;\beta )\nonumber\right.\\&{}\left.-16 G(-i,-i;\beta )+16 G(-i,i;\beta )+16 G(i,-i;\beta )-16 G(i,i;\beta )\nonumber\right.\\&{}\left.+32
   G(-1,-i,-i;\beta )-32 G(-1,-i,i;\beta )-32 G(-1,i,-i;\beta )+32 G(-1,i,i;\beta )\nonumber\right.\\&{}\left.+16
   G(0,-i,-i;\beta )-16 G(0,-i,i;\beta )-16 G(0,i,-i;\beta )+16 G(0,i,i;\beta )\nonumber\right.\\&{}\left.+16 G(-i,0,-i;\beta
   )-16 G(-i,0,i;\beta )-64 G(-i,-i,-i;\beta )+64 G(-i,-i,i;\beta )\nonumber\right.\\&{}\left.+32 G(-i,i,-i;\beta )-32
   G(-i,i,i;\beta )-16 G(i,0,-i;\beta )+16 G(i,0,i;\beta )\nonumber\right.\\&{}\left.-32 G(i,-i,-i;\beta )+32 G(i,-i,i;\beta
   )+64 G(i,i,-i;\beta )-64 G(i,i,i;\beta )\nonumber\right.\\&{}\left.+32 G(1,-i,-i;\beta )-32 G(1,-i,i;\beta )-32
   G(1,i,-i;\beta )+32 G(1,i,i;\beta )\nonumber\right.\\&{}\left.+21 \zeta_3-2 \pi ^2+10 \pi ^2 l_2\right)+\mathcal{O}\left(\ep ^2\right), \\
    
    \tilde{B}_{18}={}&\frac{2}{3} i \ep (\pi ^2 (G(-i;\beta )-G(i;\beta ))+6 (G(-i,-i,-i;\beta )-G(-i,-i,i;\beta )\nonumber\\&{}-G(-i,i,-i;\beta
   )+G(-i,i,i;\beta )-G(i,-i,-i;\beta )+G(i,-i,i;\beta )\nonumber\\&{}+G(i,i,-i;\beta )-G(i,i,i;\beta )))+\mathcal{O}(\ep^2).
\end{align}
The integrals above are expressed in terms of Goncharov polylogarithms, which are generalisations of HPLs  and defined recursively as~\cite{poincare1884groupes,goncharov2010classical,goncharov2011multiple}
\begin{equation}
    G(b,a_1\cdots a_n;x)=\int_0^x\frac{dy}{y-b}G(a_1\cdots a_n;y),\quad G(;x)=1,
\end{equation}
with 
\begin{equation}
    G(0\cdots 0;x)=\frac{1}{n!}\ln^n{x}
\end{equation}
and the integration path being a straight line from 0 to $x\in\mathbb{C}$.
\subsubsection*{Heavy propagator:}
\begin{align}
    C_1={}&\frac{1}{\ep ^2}+\frac{2}{\ep }+\left(3+\frac{\pi ^2}{6}\right)+\frac{1}{3} \left(-2 \zeta_3+12+\pi ^2\right) \ep +\frac{1}{360}
   \left(-480 \zeta_3+1800+180 \pi ^2\nonumber\right.\\&{}\left.+7 \pi ^4\right) \ep ^2+\mathcal{O}(\ep ^3,w^2) , \\
    C_2={}& \frac{3}{2 \ep ^2}+\frac{9}{2 \ep }+\frac{1}{4} \left(8 \sqrt{3} \Im\left(\text{Li}_2\left(\frac{1}{2}-\frac{i
   \sqrt{3}}{2}\right)\right)+42+\pi ^2\right)+\ep  a_{1,1}+\ep ^2 a_{1,2}+\mathcal{O}(\ep ^3,w^2), \\
    C_3={}& \frac{2 w}{\ep ^2}+\frac{2 w-2 \pi }{\ep }+\left(\frac{1}{3} \left(6+\pi ^2\right) w+\pi  (4l_2-6)\right)+\ep  \left(\frac{1}{3} w
   \left(-4 \zeta_3+6+\pi ^2\right)\nonumber\right.\\&{}\left.-\frac{2}{3} \pi  \left(21+\pi ^2+6 l_2^2-18 l_2\right)\right)+\ep ^2 \left(\frac{1}{180} w
   \left(-240 \zeta_3+360+60 \pi ^2+7 \pi ^4\right)\nonumber\right.\\&{}\left.+\frac{2}{3} \pi  \left(8 \zeta_3-45+4 l_2^3-18 l_2^2+42 l_2+\pi ^2 (2l_2-3)\right)\right)+\mathcal{O}(\ep ^3,w^2), \\
   
    C_4={}& \frac{2 w}{\ep ^2}+\frac{4 w}{\ep }+\frac{1}{3} \left(24+\pi ^2\right) w+\ep  \left(\frac{2}{3} w \left(-2 \zeta_3+24+\pi
   ^2\right)+\frac{32 \pi ^2}{3}\right)+\ep ^2 \left(\frac{1}{180} w \left(-480 \zeta_3\nonumber\right.\right.\\&{}\left.\left.+5760+240 \pi ^2+7 \pi ^4\right)+\frac{64}{9} \pi ^2
   (11-12 l_2)\right)+\mathcal{O}(\ep ^3,w^2), \\
   
    C_5={}&-\frac{2}{\ep ^2}-\frac{4}{\ep }+\left(-4 \pi ^2 w-\frac{\pi ^2}{3}-8\right)+\ep  \left(16 \pi ^2 w (2l_2-1)-\frac{2}{3} \left(-2
   \zeta_3+24+\pi ^2\right)\right)\nonumber\\&{}+\ep ^2 \left(\frac{1}{180} \left(480 \zeta_3-5760-240 \pi ^2-7 \pi ^4\right)-\frac{2}{3} w \left(7 \pi ^4+96
   \pi ^2 \left(1+2 l_2^2-2l_2\right)\right)\right)\nonumber\\&{}+\mathcal{O}(\ep ^3,w^2), \\
   
    C_6={}&\frac{w}{\ep ^2}+\frac{2 w-2 \pi }{\ep }+\left(-\frac{4}{3} \pi  \left(6+\sqrt{3} \pi -3l_2\right)+\frac{1}{6} w \left(24 \sqrt{3}
   \Im\left(\text{Li}_2\left(\frac{1}{2}-\frac{i \sqrt{3}}{2}\right)\right)+24 \nonumber\right.\right.\\&{}\left.\left. +\pi ^2\right)\right)+\ep  \left(a_{2,1}-\frac{1}{6} w \left(-12
   a_{1,1}+24 \sqrt{3} \Im\left(\text{Li}_2\left(\frac{1}{2}-\frac{i \sqrt{3}}{2}\right)\right)-8 \zeta_3+7 \pi ^2+222\right)\right) \nonumber\\&{}+\ep ^2
   \left(\frac{1}{180} w \left(-360 a_{1,1}+360 a_{1,2}+480 \zeta_3-7 \pi ^4-240 \pi ^2-5760\right)+a_{2,2}\right)+\mathcal{O}(\ep ^3,w^2), \\
    C_7={}&2 \pi ^2+\ep  \left(16 \pi ^2 w-8 \pi ^2 (l_2-1)\right)+\ep ^2 \left(32 \pi ^2 w (3-4 l_2)+\pi ^4+8 \pi ^2 \left(3+2 l_2^2-4
   l_2\right)\right)\nonumber\\&{}+\mathcal{O}(\ep ^3,w^2), \\
    C_8={}&-\frac{8 (\pi  w)}{\ep }+\left(4 \pi  w (4l_2-4)+4 \pi ^2\right)+\ep  \left(-\frac{4}{3} \pi  w \left(24+2 \pi ^2+12 l_2^2-24l_2\right)\nonumber\right.\\&{}\left.-16 \pi ^2 (l_2-1)\right)+\ep ^2 \left(\frac{8}{3} w \left(4 \pi  \left(2 \zeta_3-6+l_2^3-3 l_2^2+6l_2\right)+\pi ^3 (2l_2-2)\right)+2 \left(\pi ^4 \nonumber\right.\right.\\&{}\left.\left.+8 \pi ^2 \left(3+2 l_2^2-4 l_2\right)\right)\right)+\mathcal{O}(\ep ^3,w^2), \\
    
    C_9={}&-\frac{2 \pi ^2}{3 \ep }+\left(\frac{4}{3} \pi ^2 (2l_3-1)-\frac{16 \pi ^2 w}{3 \sqrt{3}}\right)+\ep  \left(\frac{4}{9} w \left(3 a_{2,1}+2
   \pi ^3+8 \left(3+\sqrt{3}\right) \pi ^2\nonumber\right.\right.\\&{}\left.\left.+\pi  \left(72+12 l_2^2-48 l_2\right)\right)+\frac{a_{3,1}}{2}\right)+\ep ^2
   \left(\frac{a_{3,2}}{2}-\frac{4}{9} w \left(6 a_{2,1}-3 a_{2,2}+8 \pi  \left(2 \zeta_3-6+l_2^3\nonumber\right.\right.\right.\\&{}\left.\left.\left.-3 l_2^2+6l_2\right)+48 \pi ^2 (4l_2-3)+2 \pi
   ^3 (2l_2-2)\right)\right)+\mathcal{O}(\ep ^3,w^2),  \\
    C_{10}={}&\frac{1}{2\ep^2}+\frac{3}{2\ep}+\frac{7}{2}+\frac{\pi^2}{4}
    +\ep\left(\frac{15}{2}+\frac{3\pi^2}{4}-\frac{4\zeta_3}{3}\right)
    +\ep^2\left(\frac{31}{2}+\frac{7\pi^2}{4}+\frac{7\pi^4}{80}-4\zeta_3 \right)\nonumber  \\&{}
    +\mathcal{O}(\ep ^3,w^2), \\    
    C_{11}={}&0+\mathcal{O}(\ep ^3,w^2), \\
    C_{12}={}&\frac{w}{\ep ^2}+\frac{2 w-2 \pi }{\ep }+\left(\frac{1}{2} \left(8+\pi ^2\right) w+4 \pi  (l_2-2)\right)+\ep  \left(w \left(-\frac{8
   \zeta_3}{3}+8+\pi ^2\right)-\frac{2}{3} \left(5 \pi ^3\nonumber\right.\right.\\&{}\left.\left.+6 \pi  \left(6+l_2^2-4 l_2\right)\right)\right)+\ep ^2 \left(\frac{1}{120} w
   \left(-640 \zeta_3+1920+240 \pi ^2+21 \pi ^4\right)+\frac{4}{3} \pi  \left(7 \zeta_3\nonumber\right.\right.\\&{}\left.\left.-48+2 l_2^3-12 l_2^2+5 \pi ^2 (l_2-2)+36 l_2)\right)\right)+\mathcal{O}(\ep ^3,w^2).
\end{align}

\section{Conclusion}
\label{scn:discussion}
We have employed the differential equations method to treat diagrams with heavy field insertions up to three-points two-loop order. Our analysis focused on vertex or form factor diagrams as these can be combined and used in studies beyond three-point order as they form the building blocks for a broad class of processes.  By including a mass scale, our results are applicable for a wider range of theories and provide an IR structure for massless models studied. 
The treatment of the heavy-heavy and heavy-light vertex diagrams was achieved explicitly by reducing integrals to $\ep$-form and expressing their results in terms of MPLs. On the other hand, the self-energy contributions needed for heavy field and residual mass renormalization were determined by a series expansion in the
off-shell heavy field energy. Exact off-shell self-energies were shown to require treatment with elliptic polylogarithms. We instead employed the Frobenius method to obtain solutions as an expansion in small off-shell energies. Thus, we provide further proof-positive of the power of the differential equations and dimensional recurrence approaches, advocating for their use even when more exotic propagators are present.

{\bf Acknowledgements:}
We are indebted to R.~N.~Lee, O.~L.~Veretin and V.~A.~Smirnov for stimulating discussions. This work was supported in part by the German Research Foundation DFG through Grant No.\ KN~365/12-1. The work of A.I.O was supported in part by the Foundation for the Advancement of Theoretical Physics and Mathematics "BASIS" and Russian Science Foundation, grant 20-12-00205.

\appendix

\section{One-loop master integrals}
\label{oneloop-masters}
Here, for completeness,  we present a calculation of one-loop master integrals.

\subsection{Heavy-heavy vertex}

In the case of heavy-heavy vertices, the single integral family for prototype diagram shown in Fig.~\ref{fig:oneloop} (a) is given by
\begin{equation}
I^{HH}_{\nu_1,\nu_2,\nu_3}=\int\frac{d^dl}{i\pi^{d/2}}\frac{1}{(l\cdot v_1)^{\nu_1}(l\cdot v_2)^{\nu_2}(l^2-M^2)^{\nu_3}} .
\end{equation}
Upon IBP reduction, all integrals in this family reduce to three master integrals $I^{HH}_{001}, I^{HH}_{011}, I^{HH}_{111}$. The first two master integrals are easy to calculate with Feynman parameters, and we have
\begin{equation}
I^{HH}_{001} = -\Gamma (\ep - 1)M^{2-2\ep}\, , \quad I^{HH}_{011} = \sqrt{\pi}\Gamma\left(-\frac{1}{2}+\ep\right)M^{1-2\ep}
\end{equation}
The third master integral can be conveniently calculated with the use of differential equation with respect to $w=v_1\cdot v_2$
\begin{equation}
\frac{d}{dw} I^{HH}_{111}(w) = \frac{w}{1-w^2}I^{HH}_{111}(w) - \frac{2(\ep-1)}{1-w^2}\frac{1}{M^2}I^{HH}_{001}\, .\label{eqn:odeHH}
\end{equation}
One can then easily solve Eq.~\eqref{eqn:odeHH} by variation of parameters and boundary condition, $I^{HH}_{111}(1)=-2M^{-2\ep}\Gamma (\ep)$, to obtain,
\begin{equation}
I^{HH}_{111} = -\frac{2\Gamma (\ep)M^{-2\ep}}{\sqrt{1-w^2}}\arccos (w)\, ,
\end{equation}
which mimics the well-known result from Feynman parametrization and its modification for HQET-like propagators~\cite{manohar2007heavy}.

\subsection{Heavy-light vertex}
Here, the single integral family for prototype diagram shown in Fig.~\ref{fig:oneloop} (b) is given by
\begin{equation}
I^{HL}_{\nu_1,\nu_2,\nu_3}=\int\frac{d^dl}{i\pi^{d/2}}\frac{1}{(l\cdot v_1)^{\nu_1}((l+p_2)^2-m^2)^{\nu_2}(l^2)^{\nu_3}} .
\end{equation}
Upon IBP reduction all integrals in this family reduce to two master integrals $I^{HL}_{010}, I^{HL}_{110}$. The first master we already have encountered in previous sub-section
\begin{equation}
I^{HL}_{010} = -\Gamma (\ep - 1)m^{2-2\ep}\, 
\end{equation}
The second master integral can be again calculated with the use of differential equation, this time  with respect to $w=v_1\cdot p_2/m$
\begin{equation}
\frac{d}{dw}I^{HL}_{111}(w) = \frac{w (1-2\ep)}{1-w^2}I^{HL}_{110}(w)+\frac{2(1-\ep)}{1-w^2}\frac{1}{m}I^{HL}_{010} \label{eqn:odeHL}
\end{equation}
The Eq.~\eqref{eqn:odeHL} is easily solved by variation of parameters and boundary condition, $I^{HL}_{110}(0)=\sqrt{\pi}\Gamma (-1/2+\ep)m^{1-2\ep}$, to obtain,
\begin{equation}
I^{HL}_{110} = -2m^{1-2\ep}\Gamma (\ep) w ~_2F_1 \left(1,\ep; \frac{3}{2}; w^2\right) + \sqrt{\pi}m^{1-2\ep}\Gamma\left(-\frac{1}{2}+\ep\right)(1-w^2)^{1/2-\ep}\, .
\end{equation}

\subsection{Heavy propagator}

In this case, the  single integral family for prototype diagram shown in Fig.~\ref{fig:oneloop} (c) is given by
\begin{equation}
I^{SE}_{\nu_1,\nu_2} = M^{d-\nu_1-2\nu_2}\int\frac{d^d l}{i\pi^{d/2}}\frac{1}{(l\cdot v-w)^{\nu_1}(l^2-1)^{\nu_2}}
\end{equation}
Here we have the same master integrals as in previous sub-section. These integrals are given by
\begin{align}
I^{SE}_{01} = -\Gamma (\ep - 1)M^{2-2\ep}
\end{align}
and 
\begin{equation}
I^{SE}_{11} = -2M^{1-2\ep}\Gamma (\ep) w ~_2F_1 \left(1,\ep; \frac{3}{2}; w^2\right) + \sqrt{\pi}M^{1-2\ep}\Gamma\left(-\frac{1}{2}+\ep\right)(1-w^2)^{1/2-\ep}\, .
\end{equation}

\bibliographystyle{JHEP}
\bibliography{refs}

\providecommand{\href}[2]{#2}\begingroup\raggedright\begin{thebibliography}{10}

\bibitem{pineda1997effective}
A.~Pineda and J.~Soto, {\it Effective field theory for ultrasoft momenta in
  {NRQCD} and {NRQED}},  {\em arXiv preprint hep-ph/9707481} (1997).

\bibitem{brambilla2000potential}
N.~Brambilla, A.~Pineda, J.~Soto, and A.~Vairo, {\it Potential {NRQCD}: An
  effective theory for heavy quarkonium},  {\em Nuclear Physics B} {\bf 566}
  (2000), no.~1-2 275--310.

\bibitem{georgi1990effective}
H.~Georgi, {\it An effective field theory for heavy quarks at low energies},
  {\em Physics Letters B} {\bf 240} (1990), no.~3-4 447--450.

\bibitem{jantzen2005two}
B.~Jantzen, J.~H. K{\"u}hn, A.~A. Penin, and V.~A. Smirnov, {\it {Two-loop
  high-energy electroweak logarithmic corrections in a spontaneously broken
  SU(2) gauge model}},  {\em Physical Review D} {\bf 72} (2005), no.~5 051301.

\bibitem{jantzen2006two}
B.~Jantzen and V.~A. Smirnov, {\it {The two-loop vector form factor in the
  Sudakov limit}},  {\em The European Physical Journal C-Particles and Fields}
  {\bf 47} (2006), no.~3 671--695.

\bibitem{chiu2008electroweak}
J.-y. Chiu, F.~Golf, R.~Kelley, and A.~V. Manohar, {\it Electroweak corrections
  to high energy processes using effective field theory},  {\em Physical Review
  D} {\bf 77} (2008), no.~5 053004.

\bibitem{chiu2008electroweak0}
J.-y. Chiu, F.~Golf, R.~Kelley, and A.~V. Manohar, {\it {Electroweak Sudakov
  corrections using effective field theory}},  {\em Physical review letters}
  {\bf 100} (2008), no.~2 021802.

\bibitem{chiu2008electroweak2}
J.-y. Chiu, R.~Kelley, and A.~V. Manohar, {\it {Electroweak corrections using
  effective field theory: Applications to the CERN LHC}},  {\em Physical Review
  D} {\bf 78} (2008), no.~7 073006.

\bibitem{assi1}
B.~Assi and B.~A. Kniehl, {\it Matching the standard model to {HQET and
  NRQCD}},  {\em arXiv preprint: hep-ph/2011.06447} (2020).

\bibitem{assi3}
B.~Assi and B.~A. Kniehl, {\it {Electroweak Form Factor in Sudakov and
  Threshold Regimes with Effective Field Theories}},  {\em arXiv preprint:
  hep-ph/2011.14933} (2020).

\bibitem{ovanesyan2015heavy}
G.~Ovanesyan, T.~R. Slatyer, and I.~W. Stewart, {\it Heavy dark matter
  annihilation from effective field theory},  {\em Physical Review Letters}
  {\bf 114} (2015), no.~21 211302.

\bibitem{beneke2019wino}
M.~Beneke, R.~Szafron, and K.~Urban, {\it {Wino potential and Sommerfeld effect
  at NLO}},  {\em arXiv preprint: hep-ph/1909.04584} (2019).

\bibitem{bauer2020low}
M.~Bauer, M.~Neubert, S.~Renner, M.~Schnubel, and A.~Thamm, {\it The low-energy
  effective theory of axions and {ALP}s},  {\em arXiv preprint:
  hep-ph/2012.12272} (2020).

\bibitem{mecaj2020effective}
B.~Mecaj and M.~Neubert, {\it Effective field theory for leptoquarks},  {\em
  arXiv preprint: hep-ph/2012.02186} (2020).

\bibitem{damgaard2019heavy}
P.~H. Damgaard, K.~Haddad, and A.~Helset, {\it Heavy black hole effective
  theory},  {\em Journal of High Energy Physics} {\bf 2019} (2019), no.~11
  1--27.

\bibitem{aoude2020shell}
R.~Aoude, K.~Haddad, and A.~Helset, {\it On-shell heavy particle effective
  theories},  {\em Journal of High Energy Physics} {\bf 2020} (2020),
  no.~2001.09164 1--44.

\bibitem{bern2021gravitational}
Z.~Bern, D.~Kosmopoulos, and A.~Zhiboedov, {\it Gravitational effective field
  theory islands, low-spin dominance, and the four-graviton amplitude},  {\em
  arXiv preprint: hep-ph/2103.12728} (2021).

\bibitem{smirnov2010three}
A.~V. Smirnov, V.~A. Smirnov, and M.~Steinhauser, {\it Three-loop static
  potential},  {\em Physical review letters} {\bf 104} (2010), no.~11 112002.

\bibitem{lee2016evaluating}
R.~N. Lee and V.~A. Smirnov, {\it Evaluating the last missing ingredient for
  the three-loop quark static potential by differential equations},  {\em
  Journal of High Energy Physics} {\bf 2016} (2016), no.~10 1--9.

\bibitem{ablinger2018heavy}
J.~Ablinger, J.~Bl{\"u}mlein, P.~Marquard, N.~Rana, and C.~Schneider, {\it
  Heavy quark form factors at three loops in the planar limit},  {\em Physics
  Letters B} {\bf 782} (2018) 528--532.

\bibitem{chiu2009factorization}
J.-y. Chiu, A.~Fuhrer, R.~Kelley, and A.~V. Manohar, {\it {Factorization
  structure of gauge theory amplitudes and application to hard scattering
  processes at the LHC}},  {\em Physical Review D} {\bf 80} (2009), no.~9
  094013.

\bibitem{chiesa2013electroweak}
M.~Chiesa, G.~Montagna, L.~Barze, M.~Moretti, O.~Nicrosini, F.~Piccinini, and
  F.~Tramontano, {\it {Electroweak Sudakov corrections to new physics searches
  at the LHC}},  {\em Physical review letters} {\bf 111} (2013), no.~12 121801.

\bibitem{beenakker2010supersymmetric}
W.~Beenakker, S.~Brensing, M.~Kr{\"a}mer, A.~Kulesza, E.~Laenen, and
  I.~Niessen, {\it Supersymmetric top and bottom squark production at hadron
  colliders},  {\em Journal of High Energy Physics} {\bf 2010} (2010), no.~8
  98.

\bibitem{ciafaloni2011weak}
P.~Ciafaloni, D.~Comelli, A.~Riotto, F.~Sala, A.~Strumia, and A.~Urbano, {\it
  Weak corrections are relevant for dark matter indirect detection},  {\em
  Journal of Cosmology and Astroparticle Physics} {\bf 2011} (2011), no.~03
  019.

\bibitem{gehrmann2010calculation}
T.~Gehrmann, E.~Glover, T.~Huber, N.~Ikizlerli, and C.~Studerus, {\it
  Calculation of the quark and gluon form factors to three loops in {QCD}},
  {\em Journal of High Energy Physics} {\bf 2010} (2010), no.~6 94.

\bibitem{von2017quark}
A.~von Manteuffel and R.~M. Schabinger, {\it Quark and gluon form factors to
  four-loop order in {QCD}: the $n_f^3$ contributions},  {\em Physical Review
  D} {\bf 95} (2017), no.~3 034030.

\bibitem{bernreuther2005two}
W.~Bernreuther, R.~Bonciani, T.~Gehrmann, R.~Heinesch, T.~Leineweber, and
  E.~Remiddi, {\it Two-loop {QCD} corrections to the heavy quark form factors:
  Anomaly contributions},  {\em Nuclear Physics B} {\bf 723} (2005), no.~1-2
  91--116.

\bibitem{blumlein2019heavy}
J.~Bl{\"u}mlein, P.~Marquard, N.~Rana, and C.~Schneider, {\it Heavy quark form
  factors at three loops}, .

\bibitem{ciafaloni2000electroweak}
P.~Ciafaloni and D.~Comelli, {\it {Electroweak Sudakov form factors and
  nonfactorizable soft QED effects at NLC energies}},  {\em Physics Letters B}
  {\bf 476} (2000), no.~1-2 49--57.

\bibitem{fadin2000resummation}
V.~S. Fadin, L.~Lipatov, A.~D. Martin, and M.~Melles, {\it Resummation of
  double logarithms in electroweak high energy processes},  {\em Physical
  Review D} {\bf 61} (2000), no.~9 094002.

\bibitem{kuhn2000summing}
J.~H. K{\"u}hn, A.~A. Penin, and V.~A. Smirnov, {\it {Summing up subleading
  Sudakov logarithms}},  {\em The European Physical Journal C-Particles and
  Fields} {\bf 17} (2000), no.~1 97--105.

\bibitem{feucht2004two}
B.~Feucht, J.~H. K{\"u}hn, A.~A. Penin, and V.~A. Smirnov, {\it {Two-loop
  Sudakov form factor in a theory with a mass gap}},  {\em Physical review
  letters} {\bf 93} (2004), no.~10 101802.

\bibitem{denner2001one}
A.~Denner and S.~Pozzorini, {\it One-loop leading logarithms in electroweak
  radiative corrections},  {\em The European Physical Journal C-Particles and
  Fields} {\bf 18} (2001), no.~3 461--480.

\bibitem{hori2000electroweak}
M.~Hori, H.~Kawamura, and J.~Kodaira, {\it {Electroweak Sudakov at two loop
  level}},  {\em Physics Letters B} {\bf 491} (2000), no.~3-4 275--279.

\bibitem{diffeqn1}
A.~V. Kotikov, {\it {Differential equations method: New technique for massive
  Feynman diagrams calculation}},  {\em Phys. Lett.} {\bf B254} (1991)
  158--164.

\bibitem{diffeqn2}
A.~V. Kotikov, {\it {New method of massive Feynman diagrams calculation}},
  {\em Mod. Phys. Lett.} {\bf A6} (1991) 677--692.

\bibitem{diffeqn3}
A.~V. Kotikov, {\it {Differential equations method: The Calculation of vertex
  type Feynman diagrams}},  {\em Phys. Lett.} {\bf B259} (1991) 314--322.

\bibitem{diffeqn4}
A.~V. Kotikov, {\it {Differential equation method: The Calculation of N point
  Feynman diagrams}},  {\em Phys. Lett.} {\bf B267} (1991) 123--127. [Erratum:
  Phys. Lett.B295,409(1992)].

\bibitem{diffeqn5}
E.~Remiddi, {\it {Differential equations for Feynman graph amplitudes}},  {\em
  Nuovo Cim.} {\bf A110} (1997) 1435--1452,
  [\href{http://xxx.lanl.gov/abs/hep-th/9711188}{{\tt hep-th/9711188}}].

\bibitem{epform1}
J.~M. Henn, {\it {Multiloop integrals in dimensional regularization made
  simple}},  {\em Phys. Rev. Lett.} {\bf 110} (2013) 251601,
  [\href{http://xxx.lanl.gov/abs/1304.1806}{{\tt arXiv:1304.1806}}].

\bibitem{epform2}
R.~N. Lee, {\it {Reducing differential equations for multiloop master
  integrals}},  {\em JHEP} {\bf 04} (2015) 108,
  [\href{http://xxx.lanl.gov/abs/1411.0911}{{\tt arXiv:1411.0911}}].

\bibitem{epform-criterium}
R.~N. Lee and A.~A. Pomeransky, {\it {Normalized Fuchsian form on Riemann
  sphere and differential equations for multiloop integrals}},
  \href{http://xxx.lanl.gov/abs/1707.0785}{{\tt arXiv:1707.0785}}.

\bibitem{goncharov2011multiple}
A.~B. Goncharov, {\it Multiple polylogarithms, cyclotomy and modular
  complexes},  {\em arXiv preprint: hep-ph/1105.2076} (2011).

\bibitem{remiddi2000harmonic}
E.~Remiddi and J.~A. Vermaseren, {\it Harmonic polylogarithms},  {\em
  International Journal of Modern Physics A} {\bf 15} (2000), no.~05 725--754.

\bibitem{epform-elliptics}
L.~Adams and S.~Weinzierl, {\it {The $\varepsilon$-form of the differential
  equations for Feynman integrals in the elliptic case}},  {\em Phys. Lett. B}
  {\bf 781} (2018) 270--278, [\href{http://xxx.lanl.gov/abs/1802.0502}{{\tt
  arXiv:1802.0502}}].

\bibitem{ep-regular-basis}
R.~N. Lee and A.~I. Onishchenko, {\it {$\epsilon$-regular basis for
  non-polylogarithmic multiloop integrals and total cross section of the
  process $e^+e^-\to 2(Q\bar Q)$}},  {\em JHEP} {\bf 12} (2019) 084,
  [\href{http://xxx.lanl.gov/abs/1909.0771}{{\tt arXiv:1909.0771}}].

\bibitem{levin1994elliptic}
A.~Levin and A.~Beilinson, {\it Elliptic polylogarithms},  in {\em Proceedings
  of Symposia in Pure Mathematics}, vol.~55, pp.~126--196, 1994.

\bibitem{brown2013multiple}
F.~C.~S. Brown and A.~Levin, {\it Multiple elliptic polylogarithms},  2013.

\bibitem{Adams:2014vja}
L.~Adams, C.~Bogner, and S.~Weinzierl, {\it {The two-loop sunrise graph in two
  space-time dimensions with arbitrary masses in terms of elliptic
  dilogarithms}},  {\em J. Math. Phys.} {\bf 55} (2014), no.~10 102301,
  [\href{http://xxx.lanl.gov/abs/1405.5640}{{\tt arXiv:1405.5640}}].

\bibitem{bloch2015elliptic}
S.~Bloch and P.~Vanhove, {\it The elliptic dilogarithm for the sunset graph},
  {\em Journal of Number Theory} {\bf 148} (2015) 328--364.

\bibitem{adams2016kite}
L.~Adams, C.~Bogner, A.~Schweitzer, and S.~Weinzierl, {\it The kite integral to
  all orders in terms of elliptic polylogarithms},  {\em Journal of
  Mathematical Physics} {\bf 57} (2016), no.~12 122302.

\bibitem{remiddi2017elliptic}
E.~Remiddi and L.~Tancredi, {\it An elliptic generalization of multiple
  polylogarithms},  {\em Nuclear Physics B} {\bf 925} (2017) 212--251.

\bibitem{broedel2018elliptic}
J.~Broedel, C.~Duhr, F.~Dulat, and L.~Tancredi, {\it Elliptic polylogarithms
  and iterated integrals on elliptic curves. part i: general formalism},  {\em
  Journal of High Energy Physics} {\bf 2018} (2018), no.~5 1--54.

\bibitem{Broedel:2018iwv}
J.~Broedel, C.~Duhr, F.~Dulat, B.~Penante, and L.~Tancredi, {\it {Elliptic
  symbol calculus: from elliptic polylogarithms to iterated integrals of
  Eisenstein series}},  {\em JHEP} {\bf 08} (2018) 014,
  [\href{http://xxx.lanl.gov/abs/1803.1025}{{\tt arXiv:1803.1025}}].

\bibitem{Broedel:2018qkq}
J.~Broedel, C.~Duhr, F.~Dulat, B.~Penante, and L.~Tancredi, {\it {Elliptic
  Feynman integrals and pure functions}},  {\em JHEP} {\bf 01} (2019) 023,
  [\href{http://xxx.lanl.gov/abs/1809.1069}{{\tt arXiv:1809.1069}}].

\bibitem{Broedel:2019tlz}
J.~Broedel and A.~Kaderli, {\it {Functional relations for elliptic
  polylogarithms}},  {\em J. Phys. A} {\bf 53} (2020), no.~24 245201,
  [\href{http://xxx.lanl.gov/abs/1906.1185}{{\tt arXiv:1906.1185}}].

\bibitem{weinzierl2021modular}
S.~Weinzierl, {\it Modular transformations of elliptic {Feynman} integrals},
  {\em Nuclear Physics B} {\bf 964} (2021) 115309.

\bibitem{bezuglov2021massive}
M.~Bezuglov, A.~Onishchenko, and O.~Veretin, {\it Massive kite diagrams with
  elliptics},  {\em Nuclear Physics B} {\bf 963} (2021) 115302.

\bibitem{bloch2015feynman}
S.~Bloch, M.~Kerr, and P.~Vanhove, {\it {A Feynman integral via higher normal
  functions}},  {\em Compositio Mathematica} {\bf 151} (2015), no.~12
  2329--2375.

\bibitem{Primo:2017ipr}
A.~Primo and L.~Tancredi, {\it {Maximal cuts and differential equations for
  Feynman integrals. An application to the three-loop massive banana graph}},
  {\em Nucl. Phys. B} {\bf 921} (2017) 316--356,
  [\href{http://xxx.lanl.gov/abs/1704.0546}{{\tt arXiv:1704.0546}}].

\bibitem{adams2018planar}
L.~Adams, E.~Chaubey, and S.~Weinzierl, {\it Planar double box integral for top
  pair production with a closed top loop to all orders in the dimensional
  regularization parameter},  {\em Physical review letters} {\bf 121} (2018),
  no.~14 142001.

\bibitem{adams2018analytic}
L.~Adams, E.~Chaubey, and S.~Weinzierl, {\it Analytic results for the planar
  double box integral relevant to top-pair production with a closed top loop},
  {\em Journal of High Energy Physics} {\bf 2018} (2018), no.~10 1--77.

\bibitem{Bourjaily:2017bsb}
J.~L. Bourjaily, A.~J. McLeod, M.~Spradlin, M.~von Hippel, and M.~Wilhelm, {\it
  {Elliptic Double-Box Integrals: Massless Scattering Amplitudes beyond
  Polylogarithms}},  {\em Phys. Rev. Lett.} {\bf 120} (2018), no.~12 121603,
  [\href{http://xxx.lanl.gov/abs/1712.0278}{{\tt arXiv:1712.0278}}].

\bibitem{Bourjaily:2018ycu}
J.~L. Bourjaily, Y.-H. He, A.~J. Mcleod, M.~Von~Hippel, and M.~Wilhelm, {\it
  {Traintracks through Calabi-Yau Manifolds: Scattering Amplitudes beyond
  Elliptic Polylogarithms}},  {\em Phys. Rev. Lett.} {\bf 121} (2018), no.~7
  071603, [\href{http://xxx.lanl.gov/abs/1805.0932}{{\tt arXiv:1805.0932}}].

\bibitem{Bourjaily:2018yfy}
J.~L. Bourjaily, A.~J. McLeod, M.~von Hippel, and M.~Wilhelm, {\it {Bounded
  Collection of Feynman Integral Calabi-Yau Geometries}},  {\em Phys. Rev.
  Lett.} {\bf 122} (2019), no.~3 031601,
  [\href{http://xxx.lanl.gov/abs/1810.0768}{{\tt arXiv:1810.0768}}].

\bibitem{Bonisch:2021yfw}
K.~B\"onisch, C.~Duhr, F.~Fischbach, A.~Klemm, and C.~Nega, {\it {Feynman
  Integrals in Dimensional Regularization and Extensions of Calabi-Yau
  Motives}},  \href{http://xxx.lanl.gov/abs/2108.0531}{{\tt arXiv:2108.0531}}.

\bibitem{manohar2007heavy}
A.~V. Manohar and M.~B. Wise, {\em Heavy quark physics}, vol.~10.
\newblock Cambridge university press, 2007.

\bibitem{IBP1}
F.~V. Tkachov, {\it {A Theorem on Analytical Calculability of Four Loop
  Renormalization Group Functions}},  {\em Phys. Lett.} {\bf 100B} (1981)
  65--68.

\bibitem{IBP2}
K.~G. Chetyrkin and F.~V. Tkachov, {\it {Integration by Parts: The Algorithm to
  Calculate beta Functions in 4 Loops}},  {\em Nucl. Phys.} {\bf B192} (1981)
  159--204.

\bibitem{Libra}
R.~N. Lee, {\it {Libra: A package for transformation of differential systems
  for multiloop integrals}},  {\em Comput. Phys. Commun.} {\bf 267} (2021)
  108058, [\href{http://xxx.lanl.gov/abs/2012.0027}{{\tt arXiv:2012.0027}}].

\bibitem{frobenius1}
M.~Barkatou, T.~Cluzeau, and C.~El~Bacha, {\it Frobenius methodfor computing
  powerseries solutions of linear higher-order differential systems},  {\em
  Proceedings of the 19th In-ternational Symposium on Mathematical Theory on
  Networks and Systems (MTNS)} (2010) 1059--1066.

\bibitem{frobenius2}
B.~A. Kniehl, A.~F. Pikelner, and O.~L. Veretin, {\it {Three-loop massive
  tadpoles and polylogarithms through weight six}},  {\em JHEP} {\bf 08} (2017)
  024, [\href{http://xxx.lanl.gov/abs/1705.0513}{{\tt arXiv:1705.0513}}].

\bibitem{frobenius3}
R.~Mueller and D.~G. \"Ozt\"urk, {\it {On the computation of finite
  bottom-quark mass effects in Higgs boson production}},  {\em JHEP} {\bf 08}
  (2016) 055, [\href{http://xxx.lanl.gov/abs/1512.0857}{{\tt
  arXiv:1512.0857}}].

\bibitem{frobenius4}
K.~Melnikov, L.~Tancredi, and C.~Wever, {\it {Two-loop $gg \to Hg$ amplitude
  mediated by a nearly massless quark}},  {\em JHEP} {\bf 11} (2016) 104,
  [\href{http://xxx.lanl.gov/abs/1610.0374}{{\tt arXiv:1610.0374}}].

\bibitem{frobenius5}
R.~N. Lee, A.~V. Smirnov, and V.~A. Smirnov, {\it {Solving differential
  equations for Feynman integrals by expansions near singular points}},  {\em
  JHEP} {\bf 03} (2018) 008, [\href{http://xxx.lanl.gov/abs/1709.0752}{{\tt
  arXiv:1709.0752}}].

\bibitem{frobenius6}
R.~N. Lee, A.~V. Smirnov, and V.~A. Smirnov, {\it {Evaluating
  \textquoteleft{}elliptic\textquoteright{} master integrals at special
  kinematic values: using differential equations and their solutions via
  expansions near singular points}},  {\em JHEP} {\bf 07} (2018) 102,
  [\href{http://xxx.lanl.gov/abs/1805.0022}{{\tt arXiv:1805.0022}}].

\bibitem{frobenius7}
B.~A. Kniehl, A.~V. Kotikov, A.~I. Onishchenko, and O.~L. Veretin, {\it
  {Two-loop diagrams in non-relativistic QCD with elliptics}},  {\em Nucl.
  Phys. B} {\bf 948} (2019) 114780,
  [\href{http://xxx.lanl.gov/abs/1907.0463}{{\tt arXiv:1907.0463}}].

\bibitem{tarasov1996connection}
O.~V. Tarasov, {\it Connection between {Feynman} integrals having different
  values of the space-time dimension},  {\em Physical Review D} {\bf 54}
  (1996), no.~10 6479.

\bibitem{dimrecSunrise}
O.~V. Tarasov, {\it {Hypergeometric representation of the two-loop equal mass
  sunrise diagram}},  {\em Phys. Lett. B} {\bf 638} (2006) 195--201,
  [\href{http://xxx.lanl.gov/abs/hep-ph/0603227}{{\tt hep-ph/0603227}}].

\bibitem{lee2010space}
R.~Lee, {\it Space-time dimensionality $\mathcal{D}$ as complex variable:
  Calculating loop integrals using dimensional recurrence relation and
  analytical properties with respect to $\mathcal{D}$},  {\em Nuclear Physics
  B} {\bf 830} (2010), no.~3 474--492.

\bibitem{lee2010calculating}
R.~Lee, {\it Calculating multiloop integrals using dimensional recurrence
  relation and d-analyticity},  {\em arXiv preprint: hep-ph/1007.2256} (2010).

\bibitem{Binoth:2000ps}
T.~Binoth and G.~Heinrich, {\it {An automatized algorithm to compute infrared
  divergent multiloop integrals}},  {\em Nucl. Phys.} {\bf B585} (2000)
  741--759, [\href{http://xxx.lanl.gov/abs/hep-ph/0004013}{{\tt
  hep-ph/0004013}}].

\bibitem{Binoth:2003ak}
T.~Binoth and G.~Heinrich, {\it {Numerical evaluation of multiloop integrals by
  sector decomposition}},  {\em Nucl. Phys.} {\bf B680} (2004) 375--388,
  [\href{http://xxx.lanl.gov/abs/hep-ph/0305234}{{\tt hep-ph/0305234}}].

\bibitem{Binoth:2004jv}
T.~Binoth and G.~Heinrich, {\it {Numerical evaluation of phase space integrals
  by sector decomposition}},  {\em Nucl. Phys.} {\bf B693} (2004) 134--148,
  [\href{http://xxx.lanl.gov/abs/hep-ph/0402265}{{\tt hep-ph/0402265}}].

\bibitem{Heinrich:2008si}
G.~Heinrich, {\it {Sector Decomposition}},  {\em Int. J. Mod. Phys.} {\bf A23}
  (2008) 1457--1486, [\href{http://xxx.lanl.gov/abs/0803.4177}{{\tt
  arXiv:0803.4177}}].

\bibitem{Bogner:2007cr}
C.~Bogner and S.~Weinzierl, {\it {Resolution of singularities for multi-loop
  integrals}},  {\em Comput. Phys. Commun.} {\bf 178} (2008) 596--610,
  [\href{http://xxx.lanl.gov/abs/0709.4092}{{\tt arXiv:0709.4092}}].

\bibitem{Bogner:2008ry}
C.~Bogner and S.~Weinzierl, {\it {Blowing up Feynman integrals}},  {\em Nucl.
  Phys. Proc. Suppl.} {\bf 183} (2008) 256--261,
  [\href{http://xxx.lanl.gov/abs/0806.4307}{{\tt arXiv:0806.4307}}].

\bibitem{Kaneko:2009qx}
T.~Kaneko and T.~Ueda, {\it {A Geometric method of sector decomposition}},
  {\em Comput. Phys. Commun.} {\bf 181} (2010) 1352--1361,
  [\href{http://xxx.lanl.gov/abs/0908.2897}{{\tt arXiv:0908.2897}}].

\bibitem{Fiesta4}
A.~V. Smirnov, {\it {FIESTA4: Optimized Feynman integral calculations with GPU
  support}},  {\em Comput. Phys. Commun.} {\bf 204} (2016) 189--199,
  [\href{http://xxx.lanl.gov/abs/1511.0361}{{\tt arXiv:1511.0361}}].

\bibitem{SummerTime}
R.~N. Lee and K.~T. Mingulov, {\it {Introducing SummerTime: a package for
  high-precision computation of sums appearing in DRA method}},  {\em Comput.
  Phys. Commun.} {\bf 203} (2016) 255--267,
  [\href{http://xxx.lanl.gov/abs/1507.0425}{{\tt arXiv:1507.0425}}].

\bibitem{poincare1884groupes}
H.~Poincar{\'e}, {\it Sur les groupes des {\'e}quations lin{\'e}aires},  {\em
  Acta mathematica} {\bf 4} (1884), no.~1 201--312.

\bibitem{goncharov2010classical}
A.~B. Goncharov, M.~Spradlin, C.~Vergu, and A.~Volovich, {\it Classical
  polylogarithms for amplitudes and wilson loops},  {\em Physical review
  letters} {\bf 105} (2010), no.~15 151605.

\end{thebibliography}\endgroup
\end{document}